\documentclass[twocolumn]{autart}    % Enable this line and disable the 
\usepackage{graphicx}          % Include this line if your 
\usepackage{amsmath,amssymb,amsfonts}
\allowdisplaybreaks% for amsmath package, allow equations to break across pages
\usepackage{mathrsfs}
\usepackage{mathtools}% Mathematical tools to use with amsmath
\usepackage{mathabx}% a set of 3 mathematical symbols font series: matha, mathb and mathx
\usepackage{verbatim}
\usepackage{acronym}
\usepackage{algorithm} % for algorithms
\usepackage{algpseudocodex} % for algorithms
\usepackage{tabularx}
\usepackage{xcolor}
\usepackage{dsfont}
\usepackage{cite}
\DeclareMathOperator*{\argmax}{\arg\!\max}

\newcommand{\mc}{\mathcal}
\newcommand{\mb}{\mathbb}
\newcommand{\mf}{\mathbf}

\def\R{\mathbb{R}}% set of real numbers
\def\Z{\mathbb{Z}}% set of integers
\def\E{\mathbb{E}}% expectation
\newcommand{\tp}{\intercal}% transpose
\newcommand{\onevector}[1][]{\mathbf{1}^{ #1 }}
\newcommand{\onematrix}[2][]{\mathbf{1}^{ #1 \times #2 }}
\newcommand{\MDP}{\mathcal{M}}
\newcommand{\states}{\mathcal{S}}% state space
\newcommand{\actions}{\mathcal{A}}% action space
\def\P{\mathcal{P}}% set of transition functions
\newcommand{\Reg}[1]{\texttt{Reg-{#1}}}

\newcommand{\EXP}[2][]{\mathbb{E}_{#1}\left[ #2 \right]}% expectation
\newcommand{\PR}[2][]{\mathbb{P}_{#1}\left( #2 \right)}% probability
\newcommand{\IND}[2][]{\mathbb{I}_{#1}\left\{ #2 \right\}}% indicator function
\newcommand{\norm}[1]{\lVert #1 \rVert}% norm 
\newcommand{\Oh}{\mathcal{O}}
\newcommand{\tl}{\tilde}
\newcommand{\wtl}{\widetilde}
\def\a{\alpha}
\def\d{\delta}
\def\D{\Delta}
\def\e{\epsilon}

\def\r{\rho}

\DeclareFontFamily{U}{mathx}{}
\DeclareFontShape{U}{mathx}{m}{n}{<-> mathx10}{}
\DeclareSymbolFont{mathx}{U}{mathx}{m}{n}
\DeclareMathAccent{\widehat}{0}{mathx}{"70}
\DeclareMathAccent{\widecheck}{0}{mathx}{"71}
\newcommand{\ucb}[1]{\widehat{#1}}
\newcommand{\lcb}[1]{\widecheck{#1}}
\newcommand{\mean}[1]{\widebar{#1}}
\providecommand{\given}{}
\DeclarePairedDelimiterX{\set}[1]{\{}{\}}{%
  \renewcommand{\given}{\nonscript\;\delimsize\vert\nonscript\;\mathopen{}} #1%
} % for the definition of a set

\makeatletter
\renewcommand*{\@opargbegintheorem}[3]{\trivlist
      \item[\hskip \labelsep{\bfseries #1\ #2}] \textbf{(#3)}\ \itshape}
\makeatother

\algrenewcommand\algorithmicrequire{\textbf{Input:}}

\begin{document}

\begin{frontmatter}
\runtitle{Online Learning for Dynamic VCG Mechanism}     % Running title for regular 

\title{Online Learning for Dynamic {V}ickrey-{C}larke-{G}roves
Mechanism in Unknown Environments}              % Title, preferably not more 

\thanks[footnoteinfo]{This work was supported by the Air Force Office of Scientific 
Research under award number FA9550-23-1-0107. Corresponding author: Vincent Leon.}

\author[Leon]{Vincent Leon}\ead{leon18@illinois.edu},             % Add the 
\author[Etesami]{S. Rasoul Etesami}\ead{etesami1@illinois.edu},               % e-mail address 
% \author[Baiae]{Publius Maro Vergilius}\ead{vergilius@culture.ir}  % (ead) as shown

\address[Leon]{Department of Industrial 
\& Enterprise Systems Engineering and 
Coordinated Science Laboratory,\\
University of Illinois Urbana-Champaign, 
Urbana, IL 61801, USA}                         % Please supply                                              
\address[Etesami]{Department of Industrial 
\& Enterprise Systems Engineering,
Department of Electrical \& Computer
Engineering,\\
and Coordinated Science Laboratory, 
University of Illinois Urbana-Champaign, 
Urbana, IL 61801, USA}                         % full addresses
% \address[Baiae]{The White House, Baiae}        % here.

                             % keyword list or with the 
                                          % help of the Automatica 
                                          % keyword wizard

\begin{abstract}                          % Abstract of not more than 200 words.
We consider the problem of online dynamic mechanism design for sequential auctions in unknown environments, where the underlying market and, thus, the bidders' values vary over time as interactions between the seller and the bidders progress. We model the sequential auctions as an infinite-horizon average-reward Markov decision process (MDP). In each round, the seller determines an allocation and sets a payment for each bidder, while each bidder receives a private reward and submits a sealed bid to the seller. The state, which represents the underlying market, evolves according to an unknown transition kernel and the seller's allocation policy without episodic resets. We first extend the Vickrey-Clarke-Groves (VCG) mechanism to sequential auctions, thereby obtaining a dynamic counterpart that preserves the desired properties: efficiency, truthfulness, and individual rationality. We then focus on the online setting and develop a reinforcement learning algorithm for the seller to learn the underlying MDP and implement a mechanism that closely resembles the dynamic VCG mechanism. We show that the learned mechanism approximately satisfies efficiency, truthfulness, and individual rationality and achieves guaranteed performance in terms of various notions of regret.
\end{abstract}

\end{frontmatter}

\section{Introduction}
\label{sec:intro}

Mechanism design is an interdisciplinary subfield that intersects game theory, computer science, and economics. Originating from Nobel Prize-winning economic theories, it is primarily concerned with developing rules and structures that influence the outcomes of strategic interactions. A prominent application of mechanism design is auction theory, where it defines the rules governing the allocation of goods among rational bidders and the corresponding payments. An effective mechanism should elicit the true values from bidders and incentivize them to bid truthfully, as self-interested bidders may misreport their values to the seller to achieve an outcome that benefits them. A key milestone in mechanism design is the development of the Vickrey-Clarke-Groves (VCG) mechanism \cite{vickrey1961counter,clarke1971multipart,groves1979efficient}, which ensures truthfulness and individual rationality while also identifying an allocation outcome that maximizes social welfare.

% Since bidders are rational and self-interested, they may misreport their values to the seller in order to achieve an outcome advantageous to themselves. 

% Examples of mechanisms for auctions include uniform price mechanism, pay-as-bid mechanism, Walrasian mechanism \cite{walras2003elements}, Myerson mechanism \cite{myerson1981optimal}, etc. 

Many mechanisms, such as the VCG mechanism, are \textit{static}, meaning they are designed for one-shot static auctions. However, in many cases, auctions occur sequentially, and bidders participate in a series of auctions. When there are multiple rounds of auctions, and both the seller (mechanism designer) and bidders are allowed to use the historical information acquired to determine a static mechanism and bids in the current round, the auctions are \textit{dynamic}, and such mechanisms are referred to as dynamic mechanisms \cite{bergemann2019dynamic}. Dynamic auctions are applied in a wide variety of real-world settings, such as the allocation of CO2 emission licenses \cite{cramton2002tradeable, branzei2023learning}, online ad allocation \cite{balseiro2015repeated, balseiro2019learning, golrezaei2019dynamic}, dynamic pricing for ride-sharing platforms \cite{chen2016dynamic}, and wireless spectrum allocation \cite{milgrom2017auction, cramton2013spectrum, khaledi2015dynamic}, among others.

% In dynamic auctions, the underlying market and thus the bidders' values may change as the interaction between the seller and the bidders progresses.

In this paper, we study dynamic mechanism design for sequential auctions in an unknown environment. The bidders participate in a sequence of auctions, and the number of interactions is large, possibly even unknown. The market evolves as the auction progresses according to some unknown transition kernel, and the bidders' values change accordingly. Since the number of auctions is large and potentially unknown, each bidder aims to maximize her average utility as the number of auctions approaches infinity. The seller, on the other hand, aims to implement a \emph{dynamic} version of the VCG mechanism that is efficient, truthful, and individually rational: the dynamic mechanism encourages the bidders to bid truthfully, remain in the auctions indefinitely, and also maximizes social welfare. As a result, we model the dynamic auction as an infinite-horizon average-reward Markov decision process (MDP), where the seller is the decision-making agent in the MDP, and each bidder has her own reward function which represents her valuation.

% In each round, the seller determines an allocation action to distribute the goods among bidders and charge each bidder a payment.
% Each bidder, upon receiving her goods, receives a possibly noisy reward which also depends on the current state. 
% The underlying evolves according to the current state, seller's allocation action and the unknown transition kernel. 

The authors of \cite{bergemann2010dynamic} extend the VCG mechanism to dynamic settings. However, deploying such a dynamic mechanism in the real world is challenging because it requires the seller to have full information about the transition kernel and the bidders to know their values accurately ahead of time. This is a strong assumption, as bidders often do not know the values of goods until they receive them, and the seller may not have an accurate estimate of the underlying dynamics in advance. Therefore, in this work, we propose an online reinforcement learning (RL) algorithm for the seller to learn the dynamic VCG mechanism through interactions with the bidders, without prior information about the MDP model. 

% Instead, in a dynamic setting, instead of asking each bidder to submit her values ahead of time, it is more reasonable for the seller to ask each bidder to provide a feedback based on the reward she receives after each round. 
% Moreover, it is also challenging for the seller to get an accurate description of the underlying dynamics ahead of time as it can be complicated and unpredictable. 

The existing works on learning for dynamic VCG mechanisms \cite{kandasamy2023vcg, lyu2022pessimism, qiu2024learning} formulate dynamic auctions as a multi-arm bandit (MAB) problem or an episodic MDP. In both cases, the market environment resets to an initial state after each round or episode, and auctions restart from that initial state. Our paper considers a more realistic and sophisticated setting where the market continues to evolve without restarting once the auction starts. The major challenges in designing a learning algorithm for the dynamic VCG mechanism in an infinite-horizon MDP setting arise from the non-stationary and non-restarting properties. First, every time the seller changes his allocation policy, the stationary distribution of the underlying Markov chain changes accordingly. The seller must then wait a period of time until the Markov chain approximately reaches the new stationary distribution. Implementation and evaluation of the new allocation policy and payments are meaningful only when the underlying Markov chain is close to its stationary distribution. Therefore, the seller cannot change his allocation policy too frequently, and the waiting time must be carefully designed. Second, since the market environment does not reset, the algorithm is more susceptible to the potential untruthful behavior of bidders. An untruthful bidder may manipulate the algorithm and steer the trajectory to a bad part of the MDP. A significant number of rounds may elapse before the algorithm becomes aware of the untruthful behavior and steers the trajectory out of the undesirable part. These two challenges do not exist in the MAB and episodic MDP settings. Additionally, we also face the same challenge as the authors in \cite{kandasamy2023vcg, lyu2022pessimism, qiu2024learning}. Fictitious policies associated with payment calculation must be evaluated throughout learning without being executed. Traditional RL algorithms only deal with the executed policy and do not offer performance guarantees for any other policy that is not executed by the algorithm.

To tackle the challenge of non-stationarity, we divide the learning process into episodes of increasing length, and the allocation policy and payments are updated only at the end of each episode. Moreover, each episode starts with a mixing phase whose length is carefully designed so that the induced Markov chain approximately reaches the stationary distribution by the end of the mixing phase. To address the challenge arising from the non-restarting property, we encourage the algorithm to explore the MDP by choosing the allocation policy from a subset of stochastic policies, which enables the MDP to be sufficiently explored. This technique also naturally addresses the third challenge, as sufficient exploration ensures guaranteed evaluation of the fictitious policies that are not executed. The main contributions of this paper are summarized as follows:
% \vspace{-0.1cm}
\begin{itemize}
    \item We extend the static VCG mechanism to the dynamic setting for an infinite-horizon average-reward MDP and fully characterize the seller's allocation policy and payments when the MDP is known (offline setting). Furthermore, we prove that the offline dynamic VCG mechanism is efficient, truthful, and individually rational. 
    \item We design an online RL algorithm for the seller to learn the offline dynamic VCG mechanism when the MDP model is unknown. We show that, with high probability, the time-averaged regret in terms of social welfare, seller's utility, and bidders' utility approaches an arbitrarily small term as the auctions proceed indefinitely. Moreover, we show that the learned mechanism asymptotically converges to a dynamic mechanism that approximately satisfies efficiency, truthfulness, and individual rationality. 
\end{itemize}

\subsection*{Related Works}

% Our work is related to the areas of dynamic mechanism design \cite{parkes2003mdp-based, bergemann2010dynamic, vohra2012dynamic, bergemann2019dynamic}, learning for mechanism design \cite{kandasamy2023vcg, lyu2022pessimism, qiu2024learning, mohri2016learning, morgenstern2016learning} and reinforcement learning in stochastic games \cite{zhao2022provably, uzzaman2020reinforcement, qin2024learning, zhang2021multi}.
% Among these fields, 

While the problem of mechanism design has been extensively studied in the literature from both static and dynamic points of view \cite{bergemann2010dynamic, parkes2003mdp-based, parkes2004approximately, friedman2003pricing, pavan2014dynamic, zhang2022incentive, gallien2006dynamic, kakade2013optimal, golrezaei2017auctions, vohra2012dynamic, bergemann2019dynamic}, these works do not involve learning. In this line of research, our work is closely related to \cite{kandasamy2023vcg}, \cite{lyu2022pessimism}, and \cite{qiu2024learning}, which study learning for dynamic VCG mechanisms.

The authors of \cite{kandasamy2023vcg} study online learning for the VCG mechanism in a stochastic MAB setting and propose an online learning algorithm that mimics the dynamic version of the VCG mechanism. However, their problem is set in a simpler environment, as there are no underlying dynamics. In \cite{lyu2022pessimism}, the authors model dynamic auctions as an episodic MDP and design an offline RL algorithm to learn the dynamic VCG mechanism from data. \cite{qiu2024learning} study an online counterpart of the model proposed in \cite{lyu2022pessimism} and devise an online RL algorithm that achieves an outcome approximating the dynamic VCG mechanism that would have been designed with prior knowledge of the model, in terms of various notions of regret, approximate efficiency, individual rationality, and truthfulness. However, these works study simpler settings than ours, where their models reset and restart after each round or episode, and the evolution of the environment is not continuous throughout the entire process. Other relevant works that study online learning from the seller's perspective include \cite{dudik2020oracle-efficient, golrezaei2023incentive-aware, morgenstern2016learning, cai2017learning, mohri2016learning, han2025optimal}. Finally, our work is related to \cite{balseiro2019contextual}, \cite{branzei2023learning}, and \cite{galgana2025learning}, which study online learning in dynamic auctions from the bidder's perspective, as well as to the works \cite{etesami2024learning} and \cite{qin2024learning}, which focus on reinforcement learning in $n$-player stochastic games for convergence to an $\epsilon$-Nash equilibrium. We refer readers to Appendix \ref{ap:related-works} for a more detailed review of related works.

%Our work is closely related to \cite{bergemann2010dynamic}, where the authors study a variant of the MDP-based dynamic setting and generalize the static VCG mechanism to an infinite-horizon discounted-reward MDP, to \cite{kandasamy2023vcg}, \cite{lyu2022pessimism}, and \cite{qiu2024learning}, which study learning for dynamic VCG mechanisms in MAB and episodic MDP settings, and to \cite{etesami2024learning} and \cite{qin2024learning}, which focus on reinforcement learning in $n$-player stochastic games for convergence to an $\epsilon$-Nash equilibrium. We refer readers to Appendix \ref{ap:related-works} for a more extended review of related works.

%%

\medskip
\noindent
\textbf{Notation:}
For any positive integer \( n \), we define \( [n] \triangleq \{1, \dots, n\} \). \( \mathbb{R}_{+} \) and \( \mathbb{Z}_{+} \) denote the sets of nonnegative real numbers and nonnegative integers, respectively. We also define \( \mathbb{Z}_{++} \) as the set of positive integers. \(\onevector[n]\) and \(\onematrix[m]{n}\) denote the $n$-dimensional vector and $m\times n$ matrix of all ones, respectively.
All comparison operators between vectors and matrices are applied elementwise. We use calligraphic and uppercase letters to denote finite sets and their cardinalities, respectively (e.g., $S = |\states|$, $A = |\actions|$).
We use \(\mathcal{O}\) to denote the asymptotic upper bound and \( \widetilde{\mathcal{O}} \) to represent \( O \) while hiding logarithmic factors. \( \mathbb{I}_{\cdot} \) is the indicator function: \( \mathbb{I}(A) = 1 \) if \( A \) occurs, and \( \mathbb{I}(A) = 0 \) otherwise. Unless stated otherwise, in the context of MDPs, the probability and expectation are taken with respect to the transition kernel \( P \) and policy \( \pi \), i.e., \( \mathbb{P}(\cdot) = \mathbb{P}(\cdot \mid P, \pi) \) and \(\EXP{\cdot} = \EXP{\cdot \mid P, \pi} \).
For $t \in \mb{Z}_{+}$, we use $s^t$ and $a^t$ to denote the state and the action in round $t$, respectively. Likewise, a superscript $t$ is added to any function to denote its realized value in round $t$.

%For any positive integer $n$, we let $[n] \triangleq \{1, \cdots, n\}$.  $\R$ and $\R_{+}$ denote the sets of all real numbers and nonnegative real numbers, respectively.  $\Z$, $\Z_{+}$ and $\Z_{++}$ denote the sets of all integers, nonnegative integers, and positive integers, respectively. $ \onevector[n] $ denotes an $n$-dimensional one vector. When the dimension is clear in the given context, the superscript of $ \onevector[n] $ is omitted.  When used for vectors, comparison operators (e.g., $<$, $\leq$, $=$, $\geq$, and $>$) are elementwise comparison operators. We use calligraphic and uppercase letters to denote finite sets and their cardinality respectively (e.g., $S = |\states|$, $A = |\actions|$). We use $\Oh$ to denote the asymptotic upper bound and $\TOh$ to represent $O$ by hiding the logarithmic factors.  For an event $A$, $\PR{A}$ denotes the probability of A. $\IND{\cdot}$ is the indicator function: $\IND{A} = 1$ if A occurs; otherwise, $\IND{A} = 0$. For a random variable $X$, $\EXP{X}$ denotes the expectation of $X$.  In the context of MDP, the probability and expectation are taken with respect to transition kernel $P$ and policy $\pi$ wherever applicable, and for brevity, the dependency on $P$ and $\pi$ are suppressed wherever applicable if there is no ambiguity, i.e, $\PR{\cdot} \triangleq \PR{\cdot | P, \pi}$, and $\EXP{\cdot} \triangleq \EXP{\cdot | P, \pi}$, unless otherwise specified.

\section{Problem Formulation}
\label{sec:model}

We study sequential sealed-bid auctions with a set $[n]$ of $n$ rational bidders and one seller, which is denoted as player $0$.\footnote{We often use he/him/his to refer to the seller and she/her/hers to refer to a bidder. Moreover, we refer to both the bidders and the seller as players.} The auctions are conducted sequentially, and each bidder's private value for the goods may change over time as the underlying market evolves. We model the sequential auction as an MDP denoted by $\MDP(\states, \actions, P, \{r_i\}_{i=0}^n)$. Associated with the MDP model are a finite state space $\states$, a finite action space $\actions$, a transition kernel $P: \states \times \actions \to \D(\states)$, and reward functions $\{r_i\}_{i=0}^n$.
The state space $\states$ describes the relevant market conditions that determine the evolution of the economic environment and the rewards that bidders can receive from the seller's allocation. A state can encode, for example, asset prices, economic indicators, market features, any common public information about the items, etc.
The action space $\actions$ represents the set of all feasible allocations that the seller can choose. Note that we do not restrict our model to any specific type of auction. Instead, it can be applied to any type of auction. We provide examples of different types of auction in Appendix \ref{ap:examples}. For each $i\in [n]$, we let $r_i: \states \times \actions \to [0,1]$ denote the reward function of bidder $i$, and $r_i(s, a)$ represents her private and independent value of the goods when the seller's allocation is $a \in \actions$ under state $s \in \states$. We let $r_0: \states \times \actions \to [0, c_{\max}]$ for some constant $c_{\max} \in \R_+$ denote the reward function of the seller. The seller's reward function represents the commission that he earns from each auction, independent of the payment collected from the bidders. Finally, the transition kernel $P$ describes the dynamics of the underlying environment. The state transition probability $P(\cdot | s, a)$ depends on the current state $s \in \states$ and the seller's allocation action $a \in \actions$, and the system transitions to the next state $s' \in \states$ with probability $P(s' | s, a)$.
We impose the following uniform ergodicity assumption on the MDP model. 
% \begin{assumption}\label{assump:ErgodicMDP}
% For any stationary allocation policy $\pi$, the induced Markov chain with transition probabilities $P^{\pi}(s'|s) = \sum_{a\in \actions} P(s' | s, a) \pi(a|s)$ is aperiodic and ergodic (in this case, the MDP is called \emph{unichain MDP}), and its mixing time is uniformly bounded above by some parameter $\tau$, i.e., 
% \begin{equation}
% \label{eq:ErgodicMDP}
%     \|(\nu - \nu')^{\tp} P^{\pi}\|_1 \leq e^{-\frac{1}{\tau}} \|\nu - \nu'\|_1 \quad \forall \nu, \nu' \in \D(\states).
% \end{equation}
% \end{assumption}
% In fact, this is a standard assumption in the literature of MDP and reinforcement learning and is essential for the seller to learn the MDP and evaluate his allocation policy and price vector when the underlying MDP model is unknown to him. 
% Furthermore, we impose the following nondegeneracy assumption on the MDP model. 
An intuition of Assumption \ref{assump:NondegenerateMDP} is that under any policy, a state is reachable from any other state in one step with strictly positive probability uniformly bounded below. 
%%
% \begin{assumption}\label{assump:NondegenerateMDP}
% There exists some $\alpha > 0$ such that $\sum_{a\in \actions} P(s'|s, a) \geq \alpha$ for all $s, s'\in \states$.
% \end{assumption}
%%
\begin{assum}\label{assump:NondegenerateMDP}
There exists some $\a > 0$ such that $ P(s' | s, a) \geq \a $ for all $ s, s' \in \states $ and $ a \in \actions $.
\end{assum}

Furthermore, let $\pi : \mathcal{S} \to \mathcal{D}(\mathcal{A})$ denote the seller's \textit{allocation policy}, where $\pi(a \mid s)$ denotes the probability that the seller chooses action $a \in \mathcal{A}$ in state $s \in \mathcal{S}$. Let $p \triangleq (p_i)_{i=1}^n : \mathcal{S} \times \mathcal{A} \to \mathbb{R}^n$ denote the seller's \textit{payment policy}, where $p_i(s, a)$ is the payment made by bidder $i$ to the seller when the seller chooses action $a$ in state $s$. Depending on whether we consider the offline or online version of the dynamic auction, the interaction between the seller and the bidders may vary slightly. Details of these protocols will be discussed later. In general, each bidder submits her bids to the seller, and the seller then generates the allocation and payment policies accordingly. We note that, unlike the Markov game setting, in our model the seller is the sole decision maker and learner who takes actions and interacts with the environment in the MDP setting. The bidders do not interact with the environment directly; instead, they influence the trajectory of the MDP indirectly through their interactions with the seller via bidding.

Next, before formally introducing the dynamic auction protocols, we define the value functions and benchmarks used to evaluate the performance of our learning algorithms. We adopt infinite-horizon average rewards for both the value functions and the benchmarks. This choice reflects the fact that the duration each bidder expects to participate in the dynamic auction may be uncertain, due to underlying dynamics that are potentially unknown to her. If the bidders have incentives to stay in the auction indefinitely rather than drop out (as we will show later), it is reasonable to consider an infinite-horizon setting and use the notion of average reward. For any initial state $s$, stationary allocation policy \footnote{An allocation policy is called stationary if the probability of choosing $a \in \actions$ at round $t$ depends only on the current state $s \in \states$ and is independent of the time $t$. Otherwise, it is non-stationary. It has been shown in \cite{altman1999book} that for infinite-horizon average-reward MDPs, the set of stationary policies is complete and dominant. Therefore, it suffices to consider only stationary allocation policies when the MDP is known and when the offline version is studied.} $\pi$, and bounded function $r: \states \times \actions \to \R_+$, we define the expected average payoff as follows: 
\begin{equation} \label{eq:ExpectedAveragePayoff}
    J(\pi; r) \triangleq \lim_{T \to \infty} \frac{1}{T} \E \left[ \sum_{t=1}^T r(s^t, a^t) \Big| s^1 = s \right].
\end{equation}
% where the expectation is taken with respect to the transition kernel $P$ and allocation policy $\pi$.
Under Assumption \ref{assump:NondegenerateMDP}, $J(\pi; r)$ exists and is independent of the initial state \cite{altman1999book}.
Let 
\(
    R = \sum_{j = 0}^{n} r_j
\). 
We define \textit{average social welfare} as follows: 
% \( w(\pi) \triangleq J(\pi; R) \).
\begin{equation}
\label{eq:AverageSocialWelfare}
    w(\pi) \triangleq J(\pi; R).
    % = \lim_{T \to \infty} \frac{1}{T} \E \left[ \sum_{t=1}^T R(s^t, a^t) \Big| s^1 = s \right].
    % & =\lim_{T \to \infty} \frac{1}{T} \E \left[ \sum_{t=1}^T \sum_{j=0}^n r_j(s^t, a^t) \Big| s^1 = s \right].
\end{equation}
Given any bounded stationary payment policy $p$, we define \textit{bidder $i$'s average utility} \(u_i(\pi, p_i)\) and \textit{seller's average utility} \(u_0(\pi, p)\) as follows: 
% \(
%     u_i(\pi, p) \triangleq J(\pi;r_i) - p_i
% \)
% and 
% \(
%     u_0(\pi, p) \triangleq J(\pi;r_0) + \sum_{i=1}^n p_i
% \), respectively.
% %
%
\begin{align}
    u_i(\pi, p_i)   & \triangleq J(\pi; r_i - p_i), \label{eq:BidderUtility} \\
    u_0(\pi, p)     & \triangleq J(\pi; r_0 + \sum_{i=1}^n p_i). \label{eq:SellerUtility}
\end{align}

We introduce two dynamic auction protocols that describe the interactions between the seller and the bidders: an \textit{offline} version when the MDP is known and an \textit{online} version when the MDP is unknown.

\subsection{Offline Dynamic Auction With Known MDP}\label{subsec:offline}

The state space $\states$ and the action space $\actions$ are public information known to the seller and the bidders. Each player $i \in [n] \cup \{0\}$ knows his/her own reward function $r_i$ before the sequential auction starts. Moreover, the seller knows the transition kernel $P$. In this setting, we study the offline problem: Each bidder $i \in [n]$ submits her bid $b_i : \states \times \actions \to [0, 1]$ (which is equivalent to an $S \times A$ matrix) to the seller before the dynamic auction starts, and the seller determines an allocation policy and a payment policy at the beginning of the dynamic auction. Once the sequential auction starts, the seller will stick to the same allocation and payment policies. Hence, the seller and the bidders are called offline seller and offline bidders, respectively. An offline bidder $i$ is \textit{truthful} if $b_i = r_i$; otherwise, she is \textit{untruthful}. An \textit{offline dynamic mechanism} (or simply, \textit{mechanism}) is a mapping that collects bids from each bidder $\{b_i\}_{i \in [n]}$ and maps it to a pair of allocation policy and payment policy $(\pi, p)$. The interactions between the seller and the bidders in the offline setting are summarized in Protocol~\ref{prot:offline}.

\begin{algorithm}[t]
\floatname{algorithm}{Protocol} % <-- CHANGED LINE
\caption{Offline Dynamic Auction} \label{prot:offline}
\begin{algorithmic}[1]
\State Each bidder $i \in [n]$ submits her bids $b_i$ to the seller.
\State The seller determines an allocation policy $\pi$ and a payment policy $p$.
\For{$t = 1, 2, \cdots$}
    \State The seller observes the state $s^t$ and chooses an allocation action $a^t \sim \pi(\cdot | s^t )$.
    \State The seller charges $p_i(s^t, a^t)$ to each bidder $i \in [n]$.
    \State The state transits to $s^{t+1}$.     % \Comment{This is a comment}
\EndFor
\end{algorithmic}
\end{algorithm}

On one hand, the seller aims to design a mechanism that maximizes the average social welfare defined in Eq.~\eqref{eq:AverageSocialWelfare} without knowing bidders' true private reward functions. On the other hand, each rational bidder hopes to maximize her own average utility defined in Eq.~\eqref{eq:BidderUtility} by possibly bidding untruthfully to deceive the mechanism and tends to drop out when her average utility becomes negative. An offline dynamic mechanism that achieves these desiderata will be presented in Section~\ref{sec:offline}.

% \begin{algorithm}[ht]
% \begina{algorithmic}
% \SetAlgorithmName{Protocol}{}{}
% \DontPrintSemicolon
% \caption{Online Dynamic Auctions} \label{prot:auction}
% \For{$t = 1, 2, \cdots$}{
%     The seller observes the state $s^t$ and determines an allocation action $ a^t \sim \pi^t( \cdot | s^t ) $ and payments $p^t$.\;
%     % The seller implements allocation according to $a^t$ and charges $p_i^t$ to each bidder $i \in [n]$.\;
%     Each bidder $i \in [n]$ receives a realized reward $r^t_{i}(s^t, a^t)$ whose expectation equals $r_i(s^t, a^t)$ independently drawn from some fixed distribution and pays the seller $p_i^t$. \;
%     The seller receives a realized reward $r^t_0(s^t, a^t)$ whose expectation equals $r_0(s^t, a^t)$ drawn from some fixed distribution and the payments $\sum_{i=1}^{n} p_i^t$.\;
%     Each bidder $i \in [n]$ submits her realized reward (not necessarily truthfully) to the seller as bid.\;
%     The state transits to $s^{t+1}$.\;
% }
% \end{algorithm}

\subsection{Online Dynamic Auction With Unknown MDP}\label{subsec:online}

In this setting, the state space $\states$ and the action space $\actions$ are the only information known to the seller and the bidders. Neither the seller nor the bidders know the transition kernel $P$. Moreover, we do not require that each player $i \in [n] \cup \{0\}$ knows his/her own reward function $r_i$ before the sequential auction starts. Rather, we consider a more difficult setting: Players do not know their private reward functions ahead of time. In each round $t$, after the seller observes the state $s^t$ and chooses an allocation action $a^t$, each player $i \in [n] \cup \{0\}$ receives a private stochastic bandit feedback $r^t_i(s^t, a^t)$, which is the realized value of his/her reward function independently drawn from some fixed unknown distribution $\mc{D}_i(s^t, a^t)$ with mean $r_i(s^t, a^t)$, not observed by other players. The players will learn about their private reward functions from possibly noisy data as the sequential auction continues. Hence, when the MDP is unknown (specifically, when the transition kernel and the reward functions are unknown), it is only meaningful to study the online problem: Bidders are allowed to submit bids in each round, and the seller implements possibly time-varying allocation and payment policies and learns a dynamic mechanism through the history. 

We consider the following two ways in which a bidder $i \in [n]$ participates in the online dynamic auction:
\begin{itemize}
    \item {\bf Participate by bids:} Bidder $i$ submits her bids $b_i$ (an $S \times A$ matrix) before the dynamic auction starts. Bidders who participate by bids know their private reward functions ahead of time and are offline (like in the offline problem). A bidder $i$ who participates by bids is \emph{truthful} if $b_i = r_i$; otherwise, she is \emph{untruthful}.

    \item {\bf Participate by rewards:} In each round $t$, bidder $i$ reports her realized reward $r^t_i(s^t, a^t)$ to the seller (not necessarily truthfully). This reported value of bidder $i$'s realized reward, denoted by $b_i^t$, will be used as her bid for the next round. Bidders who participate by rewards do not necessarily know their private reward functions ahead of time. A bidder $i$ who participates by rewards is \emph{truthful} if $b_i^t = r_i^t(s^t, a^t)$ for all $t$; otherwise, she is \emph{untruthful}. A bidder $i$ adopts a \emph{stationary} bidding strategy if in round $t$, she reports a sample drawn from some fixed distribution $\mc{D}_i'(s^t, a^t)$ depending only on the state $s^t$ and the seller's allocation action $a^t$; otherwise, she adopts a \emph{non-stationary} bidding strategy. It is obvious that a truthful bidding strategy is always stationary, whereas an untruthful bidder can adopt stationary or non-stationary bidding strategies.         
\end{itemize}
We allow a mixture of bidders who participate by bids and by rewards. In fact, participation by bids can be treated as a special case of participation by rewards. A bidder $i$ who participates by bids can participate by rewards by reporting the same value $b_i(s, a)$ every time the state-action pair $(s, a)$ is visited. Technically, participation by bids is simpler than, if not equivalent to, participation by rewards by reporting the $(s^t, a^t)$-th entry of $b_i$ in each round for algorithm design and analysis. This is because, once the seller knows that bidder $i$ participates by bids, he can eliminate the uncertainty for the estimation of bidder $i$'s reported reward function, which will only improve the learning performance. Therefore, in the remainder of the paper, we focus on a more challenging setting and assume that all bidders participate based on rewards. The seller learns the mechanism and, in each round $t$, determines a pair of possibly time-varying allocation and payment policies $(\pi^t, p^t)$. An \textit{online dynamic mechanism} is a mapping that takes as input the history up to the previous round, $\left\{ \left(s^{\tau}, a^{\tau}, r_0^{\tau}, (b_i^{\tau})_{i=1}^n \right) \right\}_{\tau=1}^{t-1}$, and outputs a pair of allocation and payment policies $(\pi^t, p^t)$. The interactions between the seller and the bidders in the online dynamic auction, when the MDP is unknown, are summarized in Protocol~\ref{prot:online}.

\begin{algorithm}[t]
\floatname{algorithm}{Protocol} % <-- CHANGED LINE
\caption{Online Dynamic Auction} \label{prot:online}
\begin{algorithmic}[1]
\State The bidders who participate by bids submit their bids $\{b_i\}$ to the seller. 
\For{$t = 1, 2, \cdots$}
    \State The seller determines an allocation policy $\pi^t$ and a payment policy $p^t$.
    \State The seller observes the state $s^t$ and chooses an allocation action $a^t \sim \pi^t(\cdot | s^t )$.
    \State The seller charges $p_i^t(s^t, a^t)$ to each bidder $i \in [n]$.
    \State Each player $i \in [n] \cup \{0\}$ receives a realized reward $r_i^t(s^t, a^t)$. 
    \State The bidders who participate by rewards report their realized rewards (not necessarily truthfully) $\{b_i^t\}$ to the seller as their bids for round $t+1$.
    \State The state transits to $s^{t+1}$.     % \Comment{This is a comment}
\EndFor
\end{algorithmic}
\end{algorithm}

Similarly to the offline version when the MDP is known, our objective is to design a learning algorithm for the seller to learn the unknown MDP and implement an online dynamic mechanism that approximately maximizes the average social welfare and incentivizes the bidders to report their realized rewards truthfully and not to drop out. Each rational bidder aims to maximizes her average utility by possibly submitting untruthful bids if she participates by bids or reporting a false realized reward in each round if she participates by rewards. A bidder who participates by rewards can be more subtly manipulative than a bidder who participates by bids because she can take advantage of the seller and maneuver on the trajectories that the seller relies on for learning the MDP and the online dynamic mechanism. An online learning algorithm for the dynamic mechanism that approximately achieves the desiderata will be presented in Section~\ref{sec:online}.

\section{Preliminaries}
\label{subsec:occ-mea}

In this section, we provide definitions and preliminary results on occupancy measures, which will later be used to develop and prove our main algorithmic results. 
% For background on the static VCG mechanism, we refer to the supplementary material.   

%\subsection{VCG Mechanism}
%\label{subsec:vcg}

%Briefly talk about VCG mechanism and Clark Pivot rule here.

%\subsection{Occupancy Measure of Infinite-horizon MDPs}

% Briefly talk about occupancy measure here.
For a unichain MDP with state space $\states$, action space $\actions$, and transition kernel $P$, the infinite-horizon occupancy measures corresponding to a given stationary policy $\pi$ is defined through the following three notions \cite{altman1999book}, namely state-action-next-state frequencies $q$, state-action frequencies $\rho$, and state frequencies (also called stationary distribution in the literature of Markov chain) $\nu$: 
for all $s, s' \in \states$ and $a \in \actions$,
%
% % \begin{subequations}
% \begin{alignat}{3}
% &   \text{state-action-next-state frequencies:} \quad 
% &&  q(s,a,s') \triangleq \lim_{T\to \infty} \frac{1}{T} \sum_{t=1}^T \PR{s^t = s, a^t = a, s^{t+1} = s'} \notag \\
% &&& \qquad \forall (s, a, s') \in \states \times \actions \times \states, \label{eq:OccupancyMeasure-q} \\
% &   \text{state-action frequencies:} \quad 
% &&  \rho(s,a) \triangleq \lim_{T\to \infty} \frac{1}{T} \sum_{t=1}^T \PR{s^t = s, a^t = a} \notag \\
% &&& \qquad \forall (s, a) \in \states \times \actions, \label{eq:OccupancyMeasure-rho} \\
% &   \text{state frequencies:
% % \footnote{In the literature of Markov chain, this form of occupancy measure $\nu$ is commonly called stationary distribution.}
% } \quad 
% &&  \nu(s) \triangleq \lim_{T\to \infty} \frac{1}{T} \sum_{t=1}^T \PR{s^t = s} \qquad \forall s \in \states, \label{eq:OccupancyMeasure-nu}
% \end{alignat}
% % \end{subequations}
%
\begin{align}
    q(s,a,s') & \triangleq \lim_{T\to \infty} \frac{1}{T} \sum_{t=1}^T \PR{s^t = s, a^t = a, s^{t+1} = s'}, \label{eq:OccupancyMeasure-q} \\
    \rho(s,a) & \triangleq \lim_{T\to \infty} \frac{1}{T} \sum_{t=1}^T \PR{s^t = s, a^t = a}, \label{eq:OccupancyMeasure-rho} \\
    \nu(s) & \triangleq \lim_{T\to \infty} \frac{1}{T} \sum_{t=1}^T \PR{s^t = s}. \label{eq:OccupancyMeasure-nu}
\end{align}
Intuitively, $q(s,a,s')$, $\rho(s,a)$, and $\nu(s)$ represent the long-term probabilities that the state-action-next-state triplet $(s, a, s')$, the state-action pair $(s, a)$, and the state $s$ are visited in the MDP with transition kernel $P$ under policy $\pi$. Alternatively, they also represent the proportion of time the system spends in $(s, a, s')$, $(s, a)$, and $s$, respectively, when policy $\pi$ is executed. The relationship between these three notions of the occupancy measure is given in the following:
% %
% \begin{alignat}{3}
% q(s, a, s') & = \rho(s, a) P(s'|s, a), \quad  &
% \rho(s, a) & = \nu(s) \pi(a|s), \label{eq:OccupancyMeasureRelations1} \\
% \rho(s, a) & = \sum_{s'\in \states} q(s, a, s'), \quad &
% \nu(s) & = \sum_{a\in \actions} \rho(s, a). \label{eq:OccupancyMeasureRelations2}
% \end{alignat}
% %
\begin{align}
q(s, a, s') & = \rho(s, a) P(s'|s, a), \label{eq:OccupancyMeasureRelations1} \\
\rho(s, a) & = \sum_{s'\in \states} q(s, a, s'), \label{eq:OccupancyMeasureRelations2} \\
\nu(s) & = \sum_{a\in \actions} \rho(s, a), \label{eq:OccupancyMeasureRelations3} \\
\rho(s, a) & = \nu(s) \pi(a|s). \label{eq:OccupancyMeasureRelations4} 
\end{align}
% %
In this work, we use these three notions of occupancy measure interchangeably. 

Let $\D$ denote the set of all valid occupancy measures $q$. We have the following characterization for $\D$ \cite{altman1999book, qin2024learning}.
\begin{prop}
$\D$ is a non-empty polytope and has the following representation: 
\begin{equation*}
    \D = \set*{
    q\in \R_+^{A S^2} \! \given \! \! 
    \begin{aligned}
        & \sum_{s\in \states} \sum_{a\in \actions} \sum_{s'\in \states} q(s,a,s') = 1, \\
        & \sum_{s'\in \states} \sum_{a\in \actions} q(s', a, s) \! = \! \! \sum_{a\in \actions} \sum_{s'\in \states} q(s,a,s') \\
        & \quad \forall s \in \states 
    \end{aligned}
    }.    
\end{equation*}
\end{prop}
The first constraint and the non-negativity constraint ensure that $q$ forms a valid probability measure. The second constraint, which is analogous to flow conservation, ensures the stationary property. 
For a given function $q:\states \times \actions \times \states \to [0, 1]$, let $\rho$ and $\nu$ be the functions induced by $q$ using Eqs.~\eqref{eq:OccupancyMeasureRelations1}--\eqref{eq:OccupancyMeasureRelations4}.
If $q$ is a valid occupancy measure, $\rho$ and $\nu$ are also valid occupancy measures. With slight abuse of notation, we say $\rho \in \D$ and $\nu \in \D$ if $\rho$ and $\nu$ can be obtained from some $q \in \D$ using Eqs.~~\eqref{eq:OccupancyMeasureRelations1}--\eqref{eq:OccupancyMeasureRelations4}.

Given the transition kernel $P$ of an MDP and policy $\pi$, the occupancy measures are uniquely defined through Eqs. \eqref{eq:OccupancyMeasure-q}--\eqref{eq:OccupancyMeasure-nu}. 
Conversely, any occupancy measure $q\in \D$ can induce some transition kernel $P$ and policy $\pi$, as shown in the following proposition \cite{rosenberg2019onlineconvex}.
\begin{prop}
Any valid occupancy measure $q\in \D$ can induce a transition kernel $P^q$ and a stationary policy $\pi^q$ defined as follows:
\footnote{If the denominator is zero, $P^q(\cdot |s, a)$ is chosen to be an arbitrary probability measure over $\states$, and $\pi^q(\cdot | s)$ is chosen to be an arbitrary probability measure over $\actions$.} for all $s, s' \in \states$ and $a \in \actions$,
\begin{align}
    P^q(s'|s, a) & = \frac{q(s, a, s')}{\sum_{x\in \states} q(s, a, x)}, \label{eq:InducedTransitionFunction} \\
    \pi^q(a|s) & = \frac{\sum_{s'\in \states} q(s, a, s')}{\sum_{a'\in \actions} \sum_{s'\in \states} q(s, a', s')}. \label{eq:InducedPolicy}
\end{align} 
% \begin{equation}
% \label{eq:InducedTransitionFunctionAndPolicy}
%     P^q(s'|s, a) = \frac{q(s, a, s')}{\sum_{x\in \states} q(s, a, x)} \quad \forall s \in \states, a\in \actions, s'\in \states, \quad 
%     \pi^q(a|s) = \frac{\sum_{s'\in \states} q(s, a, s')}{\sum_{a'\in \actions} \sum_{s'\in \states} q(s, a', s')} \quad \forall s\in \states, a\in \actions. 
% \end{equation} 
\end{prop}

Given a transition kernel $P$, let $\D(P)$ denote the set of valid occupancy measures whose induced transition kernels equal $P$. It is clear that $\D(P) \subseteq \D$. The following proposition provides a characterization of $\D(P)$, which is also a polytope \cite{altman1999book, qin2024learning}.
\begin{prop}
$\D(P)$ is a non-empty polytope and has the following representation: 
\begin{equation*}
% \label{eq:OccupancyMeasurePolytopeOverTransitionFunction}
\begin{aligned}
    & \D(P) = \D \cap \\
    & \set*{
        q \in \R_+^{A S^2} \given
        \begin{aligned}
            & q(s,a,s') = P(s'|s,a) \sum_{x \in \states} q(s,a,x) \\
            & \quad \forall s, s' \in \states, a \in \actions
        \end{aligned}
    }.
\end{aligned}
\end{equation*}
\end{prop}

Let $\P$ denote a collection of transition kernels, and let $\D(\P) \triangleq \bigcup_{P \in \P} \D(P)$ represent the set of occupancy measures whose induced transition kernels belong to $\P$. In general, $\D(\P)$ is not necessarily a polytope or a convex set. However, if $\P$ is carefully defined, $\D(\P)$ will have a simple characterization and desirable properties.

% \subsection{A Dual Formulation}

The introduction of occupancy measure allows us to reduce the task of evaluating a policy to that of evaluating an occupancy measure. Similarly, it also reduces the task of learning the optimal policy to that of learning the optimal occupancy measure. 
Given a unichain MDP with transition kernel $P$ and a stationary policy $\pi$, let $q^{P, \pi}$ and $\rho^{P, \pi}$ denote the two notions of occupancy measure induced by $P$ and $\pi$, which are defined by  \eqref{eq:OccupancyMeasure-q} and \eqref{eq:OccupancyMeasure-rho}, respectively. 
With slight abuse of notation, let us define $r(s,a,s') = r(s, a)$ for all $(s, a, s')\in \states \times \actions \times \states$ for any bounded reward function $r:\states \times \actions \to \R_+$. It has been shown in \cite{altman1999book} that under Assumption \ref{assump:NondegenerateMDP}, 
\begin{equation*}
    J(\pi; r) = \langle q^{P, \pi}, r \rangle = \langle \rho^{P, \pi}, r \rangle,
\end{equation*}
where $J(\pi; r)$ is the expected average payoff defined by Eq. \eqref{eq:ExpectedAveragePayoff}. Therefore, finding the optimal policy $\pi^*$ is equivalent to solving the following LP: 
\begin{equation*}
    q^* \in \argmax_{q \in \D(P)} \ \langle q, r \rangle.
\end{equation*}
The optimal policy $\pi^*$ can be induced from $q^*$ using Eq.~\eqref{eq:InducedPolicy}.

% \subsection{Shrunk Occupancy Measure Polytope}

Finally, we consider a subset of the occupancy measure polytope, namely \emph{shrunk polytope}, as defined below \cite{qin2024learning}.  
\begin{defn}[Shrunk Polytope]
\label{def:ShrunkOccupancyMeasurePolytope}
For any $\delta \in (0,1)$, the shrunk polytope of all feasible occupancy measures is defined as follows: 
\begin{equation}
\label{eq:ShrunkOccupancyMeasurePolytope}
\begin{aligned}
    & \D_{\d} \triangleq \D \cap \\
    & \set*{
        q\in \R^{A S^2}_{+} \given 
        \sum_{s'\in \states} q(s, a, s') \geq \delta 
        \quad \forall s\in \states, a\in \actions
    }.
\end{aligned}
\end{equation}
Moreover, for a given transition kernel $P$ and a collection of transition kernels $\P$, we define $\D_{\d}(P)$ and $\D_\d(\P)$ in a similar way by replacing $\D$ in Eq. \eqref{eq:ShrunkOccupancyMeasurePolytope} with $\D(P)$ and $\D(\P)$, respectively.
\end{defn}

Intuitively, we obtain $\D_\d$ from $\D$ by ``chopping off'' the extreme points of the polytope, which correspond to deterministic policies. Eq.~\eqref{eq:InducedPolicy} and Eq.~\eqref{eq:ShrunkOccupancyMeasurePolytope} imply that for any $q \in \D_\d$, \(\pi^q(a|s) \geq \delta\) for all $s \in \states$ and $a \in \actions$, meaning that any action will be chosen with probability at least $\delta$ in any state under any stationary distribution in $\D_{\d}$. Hence, by restricting a learning algorithm to $\D_\d$ instead of $\D$, we encourage the algorithm to explore during learning. Moreover, when the occupancy measure is restricted to $\D_\d$, the denominators in Eqs.~\eqref{eq:InducedTransitionFunction}--\eqref{eq:InducedPolicy} are essentially non-zero, implying that the induced transition kernel $P^q$ is unique, which is a favorable property for both the algorithm and the analysis. The following lemma states that shrinking the polytope by a sufficiently small size $\d$ will result in only a negligible loss in the payoff. Moreover, given a desired level of payoff loss, the amount of shrinking $\delta$ can be computed efficiently .

\begin{prop}[Lemma 4.3 of \cite{etesami2024learning}]
\label{prop:ShrunkPolytope}
For any $\epsilon\!>\! 0$, there exists a $\d \in (0, 1)$ that can be computed in polynomial time such that
\begin{equation*}
    \max_{q\in \D_\d(P)} \langle q, r\rangle \geq \max_{q\in \D(P)} \langle q, r\rangle - \epsilon.
\end{equation*}
\end{prop}

\section{Offline Infinite-Horizon VCG Mechanism}
\label{sec:offline}

In this section, we consider offline dynamic auction when the MDP is known and study the offline problem defined in Section~\ref{subsec:offline}. Recall that the state space $\states$ and the action space $\actions$ are public information known to the seller and the bidders, and each player knows his/her own private reward function. All bidders submit their bids to the seller ahead of time. The seller determines stationary allocation and payment policies using the bids submitted by each bidder before the dynamic auction starts, and an offline dynamic mechanism unfolds over the sequence of auctions as the seller implements the pre-determined allocation and payment policies in each round.
% The goal of the seller is to design a single allocation policy and price vector to maximize a certain benchmark. Before the entire sequence of auctions begins, each bidder independently submits a sealed bid to the seller, which is a report (not necessarily truthful) of her  $S \times A$ reward table. It is important to note that in the offline dynamic mechanism setting, all bids are submitted at the beginning, and they are used by the seller to determine an allocation policy and a price vector a priori for the entire sequence of auctions. The mechanism then unfolds over the sequence of auctions by implementing that allocation policy and the price vector over the rounds, and each bidder receives her expected average utility at the end. 
Suppose that all bidders submit their reward functions truthfully as bids to the seller. We extend the classical VCG mechanism, originally developed for single-stage auctions, to our offline sequential auction modeled by an infinite-horizon average-reward MDP.
Let 
\(
    R = \sum_{j = 0}^{n} r_j
\)
and
\(
    R_{-i} = \sum_{\substack{j = 0 \\ j\neq i}}^n r_j 
\)
for all $i \in [n]$. Denote by $\Pi$ the set of all stationary allocation policies. 
The infinite-horizon VCG mechanism is defined as follows:

\begin{defn}[Infinite-horizon VCG mechanism]
\label{def:InfiniteHorizonMDPVCGMechanism}
An offline dynamic mechanism is an infinite-horizon VCG mechanism if it receives the bids and outputs a pair $(\pi^*, p^*)$ of allocation policy and price policy that satisfy the following equations: 
% \begin{align}
%     \pi^* & \in \argmax_{\pi \in \Pi} J(\pi; R), \label{eq:DynamicVCG1} \\
%     p_i^* & = J(\pi^{*}_{-i}; R_{-i}) - J(\pi^{*}; R_{-i}), \label{eq:DynamicVCG2} \\
%     \text{where} \quad 
%     \pi^*_{-i} & \in \argmax_{\pi \in \Pi} J(\pi; R_{-i}) \quad \forall i \in [n]. \label{eq:DynamicVCG3}
% \end{align}
\begin{align}
% \label{eq:DynamicVCG}
    & \pi^* \in \argmax_{\pi \in \Pi} J(\pi; R), \label{eq:DynamicVCG-allocation} \\
    & \pi^*_{-i} \in \argmax_{\pi \in \Pi} J(\pi; R_{-i}) \quad \forall i \in [n], \label{eq:DynamicVCG-price1} \\
    & 
    \begin{aligned}
        p_i^*(s, a) = \ & J(\pi^{*}_{-i}; R_{-i}) - R_{-i}(s, a) \\
        & \quad \forall i \in [n], s \in \states, a \in \actions.
    \end{aligned} \label{eq:DynamicVCG-price2} 
\end{align}
\end{defn}

\begin{rem}
\label{rem:DynamicVCGEquivalentForm}
The pair $(\pi^*, p^*)$ can be written in terms of occupancy measures as follows: \footnote{We can also replace $q$ in Eqs.~\eqref{eq:DynamicVCGEquivalent-allocation}--\eqref{eq:DynamicVCGEquivalent-price2} with $\r$ to get occupancy measures in $\states \times \actions$.
}
\begin{align}
    & q^* \in \argmax_{q \in \D(P)} \ \langle q, R \rangle, \label{eq:DynamicVCGEquivalent-allocation} \\
    & q^*_{-i} \in \argmax_{q \in \D(P)} \ \langle q, R_{-i} \rangle \quad \forall i \in [n], \label{eq:DynamicVCGEquivalent-price1} \\
    & 
    \begin{aligned}
        p_i^*(s, a) = \ & \langle q^*_{-i}, R_{-i} \rangle - R_{-i}(s, a) \\
        & \quad \forall i \in [n], s \in \states, a \in \actions.
    \end{aligned}
     \label{eq:DynamicVCGEquivalent-price2} 
\end{align}
% \begin{equation}
% \label{eq:DynamicVCGEquivalent}
%     q^* \in \argmax_{q \in \D(P)} \ \langle q, R \rangle, \quad
%     p_i^* = \langle q^*_{-i} - q^*, R_{-i} \rangle, \
%     \text{where} \
%     q^*_{-i} \in \argmax_{q \in \D(P)} \ \langle q, R_{-i} \rangle \ \forall i \in [n]. 
% \end{equation}
%
Therefore, the offline infinite-horizon VCG mechanism can be obtained by solving $(n+1)$ LPs. 
\end{rem}

In general, a bidder may not be truthful and may submit a bid $b_i \neq r_i$ to the seller. In this case, the seller computes $\pi^*$ and $p^*$ using the equations above, replacing $r_i$ with $b_i$. Like the classical VCG mechanism, the infinite-horizon VCG mechanism satisfies the three desiderata of mechanism design: efficiency, truthfulness, and individual rationality. We first extend these definitions to the infinite-horizon dynamic setting.
\begin{defn}[Efficiency]
\label{def:Efficiency}
An offline dynamic mechanism is \emph{efficient} if the mechanism maximizes the average social welfare $w$ when all bidders are truthful. 
\end{defn}
\begin{defn}[Truthfulness]
\label{def:Truthfulness}
An offline dynamic mechanism is \emph{truthful} if for any bidder $i \in [n]$, regardless of the behavior of other bidders, her average utility $u_i$ is maximized when she is truthful, i.e., $b_i = r_i$. 
\end{defn}
\begin{defn}[Individual rationality]
\label{def:IndividualRationality}
An offline dynamic mechanism is \emph{individually rational} if for any bidder $i \in [n]$, regardless of the behavior of other bidders, her average utility $u_i$ is nonnegative when she is truthful. 
\end{defn}
Note that these three desiderata are closely related. A truthful mechanism ensures that the bidders bid truthfully, which in turn enables efficiency. Moreover, if the mechanism is individually rational, the bidders have incentives to stay in the sequential auction and continue playing the dynamic game with the seller indefinitely rather than drop out. This justifies our use of the notion of average reward function. The following lemma stipulates that our infinite-horizon VCG mechanism indeed satisfies the three desiderata defined above. 
\begin{thm}\label{thm:InfiniteHorizonMDPVCGMechanism}
The infinite-horizon VCG mechanism is efficient, truthful, and individually rational. 
\end{thm}
The proof of Theorem \ref{thm:InfiniteHorizonMDPVCGMechanism} follows the dynamic pivot mechanism in \cite{bergemann2010dynamic} and is similar to the Markov VCG mechanism in \cite{qiu2024learning}. For completeness, we provide the proof in Appendix \ref{ap:offline}.

\section{Online Learning for the Infinite-Horizon VCG Mechanism with Unknown MDP} 
\label{sec:online}

Deriving the infinite-horizon VCG mechanism requires the seller to have complete information about the MDP model $\MDP(\states, \actions, P, \{r_i\}_{i=0}^n)$, which, in many scenarios, is unrealistic. A natural question is whether the seller could learn the MDP model---particularly the transition kernel $P$ and the reward functions $\{r_i\}_{i=0}^n$---and mimic the mechanism through interactions with the environment and the bidders. 
In this section, we consider the more challenging setting when the MDP is unknown and study the online dynamic auction defined in Section \ref{subsec:online}. Recall that in this setting, the only information known to the seller and the bidders is the state space $\states$ and the action space $\actions$, and they are public information. We do not require that each player $i\in [n] \cup \{0\}$ knows his/her own reward function $r_i$ before the sequential auction starts, and we allow players to learn their reward functions by receiving stochastic bandit feedback from the environment The interactions between the seller and the bidders in the online dynamic auction when the MDP is unknown are summarized in Protocol \ref{prot:online}.

We develop an online reinforcement learning algorithm for the seller to learn and mimic the offline infinite-horizon VCG mechanism defined in Section \ref{sec:offline}. The objective is to ensure that the learned online dynamic mechanism asymptotically resembles the offline infinite-horizon VCG mechanism and approximately satisfies the desiderata---efficiency, truthfulness, and individual rationality.
To evaluate the performance of our devised learning algorithm, we first compare the learned mechanism to the benchmark---the offline infinite-horizon VCG mechanism---which outputs a pair $(\pi^{*}, p^{*})$ of allocation policy and payments, as defined by Eqs.~\eqref{eq:DynamicVCG-allocation}--\eqref{eq:DynamicVCG-price2}. Let 
\(
    R^{t} = \sum_{i=0}^{n} r_{i}^{t}
\),
\(
    u_{0}^{t} = r_{0}^{t} + \sum_{i=1}^{n} p_{i}^{t}
\), and 
\(
    u_{i}^{t} = r_{i}^{t} - p_{i}^{t}
\) for all \(i \in [n]\).
We use the following notions to evaluate the algorithm's performance with respect to the benchmark, namely social welfare regret $\Reg{SW}(T)$, seller's regret $\Reg{SELL}(T)$, and bidders' regret $\Reg{BID}(T)$:
%
% \begin{align}
%     \text{Social welfare regret: } \ & \Reg{SW}(T) \triangleq T w(\pi^{*}) - \sum_{t=1}^{T} R^{t} \\
%     \text{Seller's regret: } \ & \Reg{SELL}(T) \triangleq T u_{0}(\pi^{*}, p^{*}) - \sum_{t=1}^{T} u_{0}^{t}, \\
%     \text{Bidders' regret: } \ & \Reg{BID}(T) \triangleq T \sum_{i=1}^{n} u_{i}(\pi^{*}, p^{*}) - \sum_{t=1}^{T} \sum_{i=1}^{n} u_{i}^{t},
% \end{align}
\begin{align*}
    \Reg{SW}(T) & \triangleq T w(\pi^{*}) - \EXP{\sum_{t=1}^{T} R^{t}}, \\
    \Reg{SELL}(T) & \triangleq T u_{0}(\pi^{*}, p^{*}) - \EXP{\sum_{t=1}^{T} u_{0}^{t}}, \\
    \Reg{BID}(T) & \triangleq T \sum_{i=1}^{n} u_{i}(\pi^{*}, p^{*}) - \EXP{\sum_{t=1}^{T} \sum_{i=1}^{n} u_{i}^{t}},
\end{align*}
where the expectation is taken with respect to the randomness of the MDP and the randomness of the algorithm itself.

% where $ w(\pi^{*}) $ and $ u_{i}(\pi^{*}, p^{*}) $ denote the average social welfare and the average utility of player $i \in [n] \cup \{0\} $ when $ (\pi^{*}, p^{*}) $ is implemented, and the expectation is taken with respect to the randomness of the MDP and the randomness of the algorithm itself.

Note that by definition, $ \Reg{SW}(T) = \Reg{SELL}(T) + \Reg{BID}(T)$.
This also indicates that, given a fixed bounded social welfare regret, the regrets of the seller and the bidders are on opposite sides: a smaller regret for one side results in a larger regret for the other side. This is intuitive, as the seller and the bidders have conflicting utilities. Hence, the regrets for both sides must be bounded in order to demonstrate the effectiveness of learning. In this paper, we present a version of the algorithm favorable to the seller that results in a smaller seller's regret and a slightly larger bidders' regret. 
An alternative version favorable to bidders is discussed in Section \ref{sec:alternative}.

Apart from the gaps with respect to the offline infinite-horizon VCG mechanism benchmark, we also aim to understand approximately how well the learned mechanism achieves the three desiderata: efficiency, truthfulness, and individual rationality. $\Reg{SW}(T)$ is an appropriate metric for efficiency, as it represents the suboptimality with respect to the maximum average social welfare. In this regard, we define the following notion of approximate efficiency.
\begin{defn}[$\e$-Approximate efficiency]
\label{def:ApproximateEfficiency}
A learning algorithm achieves \emph{$\epsilon$-approximate efficiency} (or is \emph{$\epsilon$-approximately efficient}) if \( \lim_{T \to \infty} \frac{1}{T} \Reg{SW}(T) \leq \e \)
when all bidders are truthful.
\end{defn}
As for truthfulness and individual rationality, the following two notions are defined to quantify how much the learned mechanism deviates from a truthful, individually rational mechanism. 
% %
% \begin{definition}[$\e$-Approximate truthfulness]
% \label{def:ApproximateTruthfulness}
% For any bidder $i \in [n]$, let $ \{ u_{i}^{t} \}_{t=1}^{T} $ and $ \{ \tilde{u}_{i}^{t}\}_{t=1}^{T} $ denote her realized utilities when she bids truthfully and untruthfully, respectively. 
% A learning algorithm achieves \emph{$\e$-approximate truthfulness} (or is \emph{$\e$-approximately truthful}) if for any bidder $ i \in [n] $, regardless of the behavior of other bidders, $ \lim_{T \to \infty} \frac{1}{T} \EXP{ \sum_{t=1}^{T} (\tilde{u}_{i}^{t} - u_{i}^{t})} \leq \e $.
% \end{definition} 
% %
%
\begin{defn}[Approximate truthfulness]
\label{def:ApproximateTruthfulness}
For any bidder $i \in [n]$, let $ \{ u_{i}^{t} \}_{t=1}^{T} $ and $ \{ \tilde{u}_{i}^{t}\}_{t=1}^{T} $ denote her realized utilities when she is truthful and untruthful, respectively. 
Suppose all bidders other than $i$ adopt stationary bidding strategies. 
A learning algorithm achieves \emph{approximate truthfulness} (or is \emph{approximately truthful}) if 
\( \lim_{T \to \infty} \frac{1}{T} \E[\sum_{t=1}^{T} (\tilde{u}_{i}^{t} - u_{i}^{t})] \leq 0 \), where the expectation is taken with respect to the randomness of the MDP and of the algorithm.
\end{defn} 
% %
% \begin{definition}[$\e$-Approximate individual rationality]
% \label{def:ApproximateIndividualRationality}
% A learning algorithm achieves \emph{$\e$-approximate individual rationality} (or is \emph{$\e$-approximately individually rational}) if for any bidder $i \in [n]$, regardless of the behavior of other bidders, $ \lim_{T \to \infty} \frac{1}{T} \EXP{ \sum_{t=1}^{T} u_{i}^{t} } \geq -\e $.
% \end{definition} 
% %
%
\begin{defn}[Approximate individual rationality]
\label{def:ApproximateIndividualRationality}
A learning algorithm achieves \emph{approximate individual rationality} (or is \emph{approximately individually rational}) if for any bidder $i \in [n]$, regardless of the behavior of other bidders, \( \lim_{T \to \infty} \frac{1}{T} \E[\sum_{t=1}^{T} u_{i}^{t}] \geq 0 \) when she is truthful, where the expectation is taken with respect to the randomness of the MDP and of the algorithm.
\end{defn} 

Our online learning algorithm, called \texttt{IHMDP-VCG}, is presented in Algorithm \ref{alg:MainAlgorithm}. 
In summary, the algorithm proceeds in episodes, and learning takes place between two consecutive episodes. We would like to emphasize that the notion of an episode in Algorithm \ref{alg:MainAlgorithm} is intrinsically distinct from that in episodic MDPs. In our algorithm, the episode is defined for the sake of presentation and analysis. 
The state continues to evolve between two consecutive episodes, whereas in episodic MDPs, the system restarts from an initial state at the beginning of each episode. 
In each episode, the algorithm maintains plausible sets of transition kernels and reward functions. 
At the end of each episode, the algorithm updates the allocation policy and payments by solving $(n+1)$ linear programs. 
We discuss the details of the algorithm in the following subsections. 
Note that in the rest of this section, we do not distinguish between the bidders' reward functions and their bids, as the seller does not know whether they bid truthfully or not. The proofs of the lemmas presented in this section are deferred to Appendix \ref{ap:online}.

%% Main Algorithm
\begin{algorithm}[ht]
\caption{Main Algorithm: \texttt{IHMDP-VCG}} 
\label{alg:MainAlgorithm}
\begin{algorithmic}[1]
\Require
    Parameters for shrunk polytope $ \e $ and $ \d $;\\
    initial occupancy measure $ \ucb{q}^{[1]} = \frac{ 1 }{ A S^{2} } \cdot \onevector[AS^{2}]$;\\
    initial payments $ \ucb{p}^{[1]} = \onevector[n]$;\\
    initial confidence set for the transition kernel $ \P^{[0]} $ given by Eq.~\eqref{eq:ConfidenceSetForTransitionKernelInitial};\\
    initial counters set to $0$.
    % $ N^{[0]}(s, a, s') = 0, N^{[0]}(s, a) = 0 \ \forall s, s' \in \states, a \in \actions $, $ \tau^{[1]} = 1 $.
\For{$ k = 1, 2, \cdots $}
    \State Compute $ \pi^{[k]} $ from $ \ucb{q}^{[k]} $ using Eq.~\eqref{eq:InducedPolicy}.
    \State \textit{Mixing phase: } 
    \For{$ t = \tau^{[k]}, \tau^{[k]} + 1, \cdots, \tau^{[k]} + d^{[k]} - 1 $}
        \State Implement allocation policy $ \pi^{[k]} $ and charge each bidder $0$.
        \State Collect reported rewards $ \{ r_{i}^{t} \}_{i=1}^{n} $ from the bidders.
    \EndFor
    \State \textit{Stationary phase: }
    \For{ $ t = \tau^{[k]} + d^{[k]}, \cdots, \tau^{[k]} + d^{[k]} + l^{[k]} - 1$ }
        \State Implement allocation policy $ \pi^{[k]} $ and charge each bidder $p_{i}^{[k]}$.
        \State Collect reported rewards $ \{ r_{i}^{t} \}_{i=1}^{n} $ from the bidders.
    \EndFor
    \State Update the counters $N^{[k]}(s, a, s')$ and $N^{[k]}(s, a)$ for all $s, s' \in \states$, $a \in \actions$.
    % , which denote the total number of visits of $(s, a, s')$ and $(s, a)$ up to the end of episode $k$, respectively.\;
    \State Update the confidence set for transition kernel $ \P^{[k]} $ using Eq.~\eqref{eq:ConfidenceSetForTransitionKernel}.
    \State Update the UCB $ \{ \ucb{r}_{i}^{k} \}_{i=0}^{n} $ and LCB $ \{ \lcb{r}_{i}^{k} \}_{i=0}^{n} $ for reward functions using Eqs.~\eqref{eq:UCBForRewardFunction}--\eqref{eq:LCBForRewardFunction}.
    \State Update occupancy measure $ \ucb{q}^{[k+1]} $ by solving the LP in Eqs.~\eqref{eq:AllocationPolicyUpdate}.
    \State Update payments $ \ucb{p}^{[k+1]} $ using Eq.~\eqref{eq:PaymentUpdate-Payment} by solving $n$ LPs given by Eq.~\eqref{eq:PaymentUpdate-Policy}.
    \State $ \tau^{[k+1]} = \tau^{[k]} + d^{[k]} + l^{[k]}$.
\EndFor
\end{algorithmic}
\end{algorithm}

\subsection{Learning in Episodes}
\label{subsec:LearningInEpisodes}
The learning process is divided into episodes of increasing length. Each episode consists of two phases: \emph{mixing phase} and \emph{stationary phase}.

%%
% Mixing phase 
% \paragraph{Mixing Phase}
\subsubsection{Mixing Phase}

Each episode $k$ starts with a mixing phase that lasts $d^{[k]}$ rounds, followed by a stationary phase. Let $\tau^{[k]}$ denote the start round of episode $k$. In the mixing phase, the seller implements the allocation policy $\pi^{[k]}$ induced by $\ucb{q}^{[k]} \in \D_{\d}$, which is calculated by the algorithm at the end of episode $(k-1)$, without charging the bidders. The purpose of the mixing phase is that, after a reasonably long period of time, the transition distribution becomes sufficiently close to the stationary distribution, making it meaningful for the seller to implement the learned mechanism and to learn the stationary distribution in the stationary phase.
Given any stationary allocation policy $\pi$, the mixing time of the induced Markov chain with transition probabilities $P^{\pi}(s'|s) = \sum_{a\in \actions} P(s' | s, a) \pi(a|s)$ is bounded above by using Assumption \ref{assump:NondegenerateMDP}, and the difference between the transition distribution and the stationary distribution decays exponentially fast, as shown in the following statement.
\begin{lem}
\label{lem:MixingTime}
    It holds that for any $\nu, \nu' \in \D(\states)$, 
    % \begin{equation}
    % \label{eq:ErgodicMDP}
    \(
        \| (\nu - \nu')^{\tp} P^{\pi} \|_{1} \leq ( 1 - \a S) \| (\nu - \nu')^{\tp} \|_{1}
    \).
    % \end{equation}
\end{lem}
%
% Let $d^{[k]}$ denote the 
% For any stationary allocation policy $\pi$, the induced Markov chain with transition probabilities $P^{\pi}(s'|s) = \sum_{a\in \actions} P(s' | s, a) \pi(a|s)$ is aperiodic and ergodic (in this case, the MDP is called \emph{unichain MDP}), and its mixing time is uniformly bounded above by some parameter $\tau$, i.e., 
% \begin{equation}
% \label{eq:ErgodicMDP}
%     \|(\nu - \nu')^{\tp} P^{\pi}\|_1 \leq e^{-\frac{1}{\tau}} \|\nu - \nu'\|_1 \quad \forall \nu, \nu' \in \D(\states).
% \end{equation}

%
% \todo{Replace $\tau$ with an expression of $\a$ throughout the manuscript. $\tau = 1 / \a S$.}

The length of the mixing phase $d^{[k]}$ must be carefully designed, as an excessively long mixing phase will result in a large seller's regret and undermine the approximate truthfulness. Moreover, an overly short mixing phase will result in non-stationarity and hence a large discrepancy between the transition distribution and the stationary distribution.
Let $\mc{F}^{[k-1]}$ be the $\sigma$-algebra adapted to all measurable events up to the end of episode $(k - 1)$.
Denote by $\nu^{t}$ and $\nu^{[k]}$ the transition distribution at round $t$ and the stationary distribution induced by $\pi^{[k]}$ in episode $k$, respectively.
For all $s \in \states$ and $a \in \actions$, let \( \r^{t}(s, a) = \nu^{t}(s) \pi^{[k]}(a|s) \) and \( \r^{[k]}(s, a) = \nu^{[k]}(s) \pi^{[k]}(a|s) \).
We provide an expression of $d^{[k]}$ in the following lemma, which ensures that the underlying Markov chain is approximately stationary during the stationary phase, without incurring large regret or excessively compromising the truthfulness of the learned mechanism.
\begin{lem}
\label{lem:ApproxStationarity}
    Let 
    \(
        d^{[k]} = \frac{\log{k}}{\a S}
    \) for all $k \in \Z_{++}$.
    Conditioned on $\mc{F}^{[k-1]}$, for any round $t$ in the stationary phase of episode $k$, it holds that 
    % \begin{equation*}
    \(
        \norm{\EXP{\r^{t}} - \r^{[k]}}_{1} = \norm{\EXP{\nu^{t}} - \nu^{[k]}}_{1} \leq  2 / \sqrt{k}
    \).
    % \end{equation*}
\end{lem}

% Later, we will provide the expression for $d^{[k]}$, which ensures that the MDP is approximately stationary during the stationary phase, without incurring large regret or excessively compromising the truthfulness of the learned mechanism.

%the distributions of $\states$, $\states \times \actions$ and $\states \times \actions \times \states$ are so close to the stationary distributions that implementation of the learned mechanism is meaningful 
%the data collected along the trajectory in this phase are meaningful and well represent the MDP under stationary distribution.

%
%% Stationary phase 
% \paragraph{Stationary Phase}
\subsubsection{Stationary Phase}

In the stationary phase, the seller continues to implement the same allocation policy $\pi^{[k]}$ as in the mixing phase. However, in each round, the seller charges $p_{i}^{[k]}$ to each bidder $i \in [n]$ and receives bids from each bidder. Since $ \ucb{q}^{[k]} \in \D_{\d} $ (see Eq.~\eqref{eq:AllocationPolicyUpdate}), and using Assumption \ref{assump:NondegenerateMDP}, one can show that in each round any $(s, a)$-pair is visited with strictly positive probability lower bounded by a constant (see Lemma \ref{lem:StrictlyPositiveProbability} in Appendix \ref{ap:AuxiliaryLemmas}).
As a result, if the stationary phase is sufficiently long, with high probability, every $(s, a)$-pair will be visited at least once in each episode.
% any $s \in \states$, $\PR{s^{t} = s | \pi^{[k]}} \geq \alpha \delta$, and for any $(s,a) \in \states \times \actions$, $\PR{(s^{t}, a^{t}) = (s, a) | \pi^{[k]}} \geq \alpha \delta^{2}$. Hence, in any round $t$, any $(s, a)$-pair is visited with strictly positive probability, lower bounded by $\alpha \delta^{2}$. 
% \todo{Change this part and make it concise, refer to the lemmas. }
This result is non-trivial, as it ensures sufficient exploration to get an accurate estimation of the MDP model and also mitigates the negative influence on seller's learning due to bidders' possibly untruthful behavior. 
Similar ideas have been used in existing works on multi-armed bandits \cite{kandasamy2023vcg} and episodic MDPs \cite{qiu2024learning} to encourage exploration. 
Let $l^{[k]}$ denote the length of the stationary phase in episode $k$.
We have the following lemma for the sufficiency of exploration.
\begin{lem}
\label{lem:SufficientExploration}
% Let $l^{[k]}$ denote the length of the stationary phase in episode $k$. 
Let
\begin{equation*}
    l^{[k]} = \frac{ \max\{4, \sqrt{k} \} \log (ASk/\zeta) }{ \a \d }
    \quad \forall k \in \Z_{++}.
\end{equation*}%
With probability at least $1 - \Oh(\zeta)$, every $(s, a)$-pair is visited at least once in each episode.
\end{lem}

% An episode ends when every $(s, a)$-pair is visited at least once during the stationary phase.
% The random number of rounds in each stationary phase ensures an accurate estimation of the MDP model and also eliminates the negative impact on model estimation due to bidders' possibly untruthful behavior.
% Similar ideas have been used in the existing works on multi-armed bandits \cite{kandasamy2020vcg} and episodic MDP \cite{lyu2022learning} to encourage exploration. 

% Therefore, the expected length of each stationary phase is bounded above by $\Oh \left(\frac{\log SA}{\alpha \delta^{2}}\right)$.

\subsection{Confidence Sets}

The algorithm maintains confidence sets for the transition kernel and each player's reward function, and updates these confidence sets at the end of each episode. With high probability, the transition kernel lies in the confidence set in each episode. Moreover, when the bidders are truthful, each player's reward function also lies in the confidence set with high probability. For any $s, s' \in \states$ and $a \in \actions$, denote by $N^{[k]}(s, a)$ and $N^{[k]}(s, a, s')$ the total number of visits to $(s, a)$ and $(s, a, s')$ up to the end of episode $k$, respectively. The empirical transition kernel in episode $k$ is defined as
\begin{equation}
\label{eq:EmpiricalTransitionKernel}
    \mean{P}^{[k]}(s'|s, a) \triangleq \frac{N^{[k]}(s, a, s')}{\max\{1, N^{[k]}(s, a)\}}, 
    % \quad \forall s, s' \in \states, a \in \actions.
\end{equation}
Furthermore, we define 
\begin{equation}
\label{eq:ConfidenceRadiusForTransitionKernel}
\begin{aligned}
    \e^{[k]}(s, a, s') \triangleq \ & 2\sqrt{\frac{\widebar{P}^{[k]}(s'|s,a) \log ({A S k}/{\zeta})}{\max\{1, N^{[k]}(s, a)-1\}}} \\
    & + \frac{14 \log ({A S k}/{\zeta})}{3 \max\{1, N^{[k]}(s, a)-1\}},
\end{aligned}
\end{equation}
which is a Bernstein-like confidence radius for updating the transition kernel confidence set defined as follows:
\begin{equation}
\label{eq:ConfidenceSetForTransitionKernel}
\begin{aligned}
    & \P^{[k]} \triangleq \P^{[k-1]} \cap \\ 
    & \set*{P \in \R_{+}^{A S^2} \given 
    \begin{aligned}
        & P(s' | s, a) - \widebar{P}^{[k]}(s' | s, a) \leq \epsilon^{[k]}(s,a,s') \\
        & \forall s, s' \in \states, a \in \actions
    \end{aligned}
     },
\end{aligned}
\end{equation}
with the initial set $\P^{[0]}$ being the set of all feasible transition kernels defined as follows: 
\begin{equation}
\label{eq:ConfidenceSetForTransitionKernelInitial}
% \begin{aligned}
    \P^{[0]} \triangleq \set*{P \in \R_{+}^{A S^2} \given
    \sum_{s' \in \states} P(s' | s, a) = 1 \ \forall s \in \states, a \in \actions}.
% \end{aligned}
\end{equation}
It has been shown that for all $k \in \Z_{++}$, the corresponding set of occupancy measures $\D(\P^{[k]})$ associated with $\P^{[k]}$ is a polytope that is characterized by a polynomial number of constraints in $S$ and $A$ \cite[Appendix A.1]{jin2020learning}.

Similarly, to estimate the reward function for each player in each episode, the algorithm computes the empirical reward function and the confidence radius for each bidder and the seller as follows: 
for all $i \in [n] \cup \{0\}$, $s \in \states$ and $a \in \actions$, 
\begin{align}
    \mean{r}_{i}^{[k]}(s, a) & \triangleq 
    % \frac{\sum_{t=1}^{\tau^{[k+1]}-1} \IND{(s^{t}, a^{t}) = (s, a)} r_{i}^{t}(s, a)}{\max \{1, N^{[k]}(s, a) \}},
    \frac{\sum_{t=1}^{\tau^{[k+1]}-1} \mb{I}^{t}\left\{(s, a)\right\} \cdot r_{i}^{t}(s, a)}{\max \{1, N^{[k]}(s, a) \}},
    \label{eq:EmpiricalRewardFunction}\\
    \beta^{[k]}(s, a) & \triangleq 
    \sqrt{\frac{2 \log (A S k n/ \zeta) }{\max \{1, N^{[k]}(s, a) \}}},
    \label{eq:ConfidenceRadiusForRewardFunction}
\end{align}
where $\mb{I}^{t}\left\{(s, a)\right\} \equiv \IND{(s^{t}, a^{t}) = (s, a)}$ is the indicator function for the event that $(s, a)$ is visited in round $t$.
The upper confidence bound (UCB) and the lower confidence bound (LCB) for reward functions are constructed as follows, respectively:
\begin{align}
    \ucb{r}_{i}^{[k]}(s, a) & \triangleq \min \left \{ c_{i}, \mean{r}_{i}^{[k]}(s, a) + c_{i} \beta_{i}^{[k]}(s, a) \right \}, \label{eq:UCBForRewardFunction} \\
    \lcb{r}_{i}^{[k]}(s, a) & \triangleq \max \left \{ 0, \mean{r}_{i}^{[k]}(s, a) - c_{i} \beta_{i}^{[k]}(s, a) \right\}, \label{eq:LCBForRewardFunction}
\end{align}
% \begin{equation}
% \label{eq:UCBAndLCBForRewardFunction}
%     \ucb{r}_{i}^{[k]}(s, a) \triangleq \min \left \{ c_{i}, \mean{r}_{i}^{[k]}(s, a) + c_{i} \beta_{i}^{[k]}(s, a) \right \}, \quad 
%     \lcb{r}_{i}^{[k]}(s, a) \triangleq \max \left \{ 0, \mean{r}_{i}^{[k]}(s, a) - c_{i} \beta_{i}^{[k]}(s, a) \right\},
% \end{equation}
%
where $c_{i} = 1$ for all $i \in [n]$, and $c_{0} = c_{\max}$, which is the upper bound of the seller's reward $r_{0}$. Given the above definitions, we have the following two lemmas. 
\begin{lem} 
\label{lem:ConfidenceSetForTransitionKernel}
With probability at least $1 - \Oh(\zeta)$, $P \in \P^{[k]}$ for all $k \in \Z_{++}$.
\end{lem}
\begin{lem}
\label{lem:ConfidenceSetForRewardFunction}
With probability at least $1 - \Oh(\zeta)$, $\lcb{r}_{i}^{[k]} \leq r_{i}\leq \ucb{r}_{i}^{[k]}$ for all $i \in [n] \cup \{0\}$ and $k \in \Z_{++}$.
% With probability at least $1 - \Oh(\zeta / n)$, $\lcb{r}_{0}^{[k]} \leq r_{0}\leq \ucb{r}_{0}^{[k]}$ for all $k \in \Z_{++}$. 
% Moreover, if bidder $i \in [n]$ is truthful, with probability at least $1 - \Oh(\zeta / n)$, $\lcb{r}_{i}^{[k]} \leq r_{i}\leq \ucb{r}_{i}^{[k]}$ for all $k \in \Z_{++}$. 
\end{lem}

\subsection{Policy Update}

The algorithm updates the allocation policy and payment vector at the end of each episode by solving $(n+1)$ linear programs in the space of occupancy measures.
With slight abuse of notation, we expand $\ucb{r}_{i}^{[k]}$ and $\lcb{r}_{i}^{[k]}$ from $\mathbb{R}_{+}^{A S}$ to $\mathbb{R}_{+}^{A S^{2}}$ by setting $\ucb{r}_{i}^{[k]}(s, a, s') = \ucb{r}_{i}^{[k]}(s, a)$ and $\lcb{r}_{i}^{[k]}(s, a, s') = \lcb{r}_{i}^{[k]}(s, a)$ for all $k \in \Z_{++}$, $i \in [n] \cup \{0\}$, $s, s' \in \states$ and $a \in \actions$.
Moreover, let \(\ucb{R}^{[k]} = \sum_{i=0}^{n} \ucb{r}_{i}^{[k]}\), \(\lcb{R}^{[k]} = \sum_{i=0}^{n} \lcb{r}_{i}^{[k]}\), \(\ucb{R}_{-i}^{[k]} = \sum_{\substack{j = 0 \\ j \neq i}}^{n} \ucb{r}_{j}^{[k]}\), and \(\lcb{R}_{-i}^{[k]} = \sum_{\substack{j = 0 \\ j \neq i}}^{n} \lcb{r}_{j}^{[k]}\).
%
% \begin{equation*}
% \begin{aligned}
%     & \ucb{R}^{[k]} = \sum_{i=0}^{n} \ucb{r}_{i}^{[k]}, \quad 
%     \lcb{R}^{[k]} = \sum_{i=0}^{n} \lcb{r}_{i}^{[k]},  \quad \\ 
%     & \ucb{R}_{-i}^{[k]} = \sum_{\substack{j = 0 \\ j \neq i}}^{n} \ucb{r}_{j}^{[k]}, \quad 
%     \lcb{R}_{-i}^{[k]} = \sum_{\substack{j = 0 \\ j \neq i}}^{n} \lcb{r}_{j}^{[k]}, \quad \forall i \in [n].
% \end{aligned}
% \end{equation*}
%

To compute the allocation policy $\pi^{[k+1]}$ for implementation in episode \( (k+1) \), the algorithm solves the following linear program over the shrunk occupancy measure polytope $\D_{\d}(\P^{[k]}) \triangleq \D_{\d} \cap \D(\P^{[k]})$ at the end of episode $k$: 
\begin{equation}
\label{eq:AllocationPolicyUpdate}
    \ucb{q}^{[k+1]} \in \argmax_{q \in \D_{\d}(\P^{[k]})} \langle q, \ucb{R}^{[k]} \rangle,
\end{equation}
and then induces the allocation policy $\pi^{[k+1]}$ from $\widehat{q}^{[k+1]}$ using Eq.~\eqref{eq:InducedPolicy}.

As suggested by Eq.~\eqref{eq:DynamicVCGEquivalent-price2} in Remark \ref{rem:DynamicVCGEquivalentForm}, the payment for each bidder \( i \in [n] \) consists of two components, and calculating the first component involves solving another LP. 
Correspondingly, in the online setting, the algorithm solves the following LP for each bidder \( i \in [n] \) to compute the first component of her payment: 
\begin{equation}
\label{eq:PaymentUpdate-Policy}
    \ucb{q}_{-i}^{[k+1]} \in \argmax_{q \in \D_{\d}(\P^{[k]})} \langle q, \ucb{R}_{-i}^{[k]}  \rangle.
\end{equation}
The payment for bidder $i \in [n]$ is hence 
\begin{equation}
\label{eq:PaymentUpdate-Payment}
    % \ucb{p}_{i}^{[k+1]} = \langle \ucb{q}_{-i}^{[k+1]}, \ucb{R}_{-i}^{[k]} \rangle - \langle \ucb{q}^{[k+1]}, \lcb{R}_{-i}^{[k]} \rangle.
    \begin{aligned}
    \ucb{p}_{i}^{[k+1]}(s, a) = \ & \langle \ucb{q}_{-i}^{[k+1]}, \ucb{R}_{-i}^{[k]} \rangle - \lcb{R}_{-i}^{[k]}(s, a) \\
    & \quad \forall s\in \states, a\in \actions.
    \end{aligned}
\end{equation}
The first component on the right-hand side of Eq.~\eqref{eq:PaymentUpdate-Payment} comes from Eq.~\eqref{eq:PaymentUpdate-Policy}.
% , and the occupancy measure appearing in the second component comes from Eq.~\eqref{eq:AllocationPolicyUpdate}. 
Note that in the second component, the LCB $\lcb{R}_{-i}^{[k]}$, rather than the UCB $\ucb{R}_{-i}^{[k]}$, is used.

Unlike conventional online learning problems, where the learner aims to maximize a single reward function and implements and evaluates the computed optimal policy that steers the trajectory toward optimizing the objective, in our problem of learning the dynamic VCG mechanism, the seller must learn and evaluate \( (n+1) \) policies. However, only \( \ucb{q}^{[k+1]} \) is implemented, while the policies \( \{ \ucb{q}_{-i}^{[k+1]} \}_{i=1}^{n} \) need to be evaluated without being implemented, based on the trajectory generated by \( \ucb{q}^{[k+1]} \), which may be manipulated by untruthful bidders. Therefore, to achieve approximate accuracy and optimality of the payments \( \{ \ucb{p}_{i}^{[k+1]} \}_{i=1}^{n} \) and to ensure that the learned mechanism is approximately truthful and individually rational, sufficient exploration during the learning process must be ensured. As mentioned before, this is achieved by choosing an occupancy measure in \( \Delta_{\delta}(\mathcal{P}^{[k]}) \) rather than \( \Delta(\mathcal{P}^{[k]}) \). This also makes the problem of learning the dynamic VCG mechanism more challenging than a standard online reinforcement learning problem.

\subsection{An Alternative Version Favorable to Bidders}
\label{sec:alternative}

The version of the learning algorithm presented in Algorithm \ref{alg:MainAlgorithm} is favorable to the seller. 
The payment \( \ucb{p}_{i}^{[k+1]} \) defined by Eq.~\eqref{eq:PaymentUpdate-Payment} is a biased estimator of \( p_{i}^{*} \).
With high probability, \( p_{i}^{*} \) is upper-bounded by \( \ucb{p}_{i}^{[k+1]} \) for each \( i \in [n] \) and \( k \in \mathbb{Z}_{++} \). 
Hence, the seller tends to overcharge each bidder by a small, bounded amount in each round of the stationary phase. Hence, this setting results in a smaller seller's regret but a larger bidders' regret. 
This version of the algorithm can be adopted when the algorithm designer favors the seller more than bidders.
% This is desirable for the seller but unfavorable for the bidders, as it results in a smaller seller's regret but a larger bidder's regret. This also makes sense, as the algorithm is designed for the seller. 

%

Alternatively, we can devise a version of the algorithm favorable to bidders as follows. 
We can design an underestimator of the payment for each bidder \( i \in [n] \) as follows:
\begin{align}
    \lcb{q}_{-i}^{[k+1]} \in & \argmax_{q \in \D_{\d}(\P^{[k]})} \langle q, \lcb{R}_{-i}^{[k]} \rangle, \label{eq:AlternativePaymentUpdate-Policy} \\
    \lcb{p}_{i}^{[k+1]}(s, a) = \ & \langle \lcb{q}_{-i}^{[k+1]}, \lcb{R}_{-i}^{[k]} \rangle - \ucb{R}_{-i}^{[k]}(s, a) \label{eq:AlternativePaymentUpdate-Payment} \\
    & \quad \forall s \in \states, a \in \actions. \notag
\end{align}
% \begin{equation}
% \label{eq:AlternativePaymentUpdate}
%     \lcb{q}_{-i}^{[k+1]} \in \argmax_{q \in \D_{\d}(\P^{[k]})} \langle q, \lcb{R}_{-i}^{[k]} \rangle, \quad
%     \lcb{p}_{i}^{[k+1]}(s, a) = \langle \lcb{q}_{-i}^{[k+1]}, \lcb{R}_{-i}^{[k]} \rangle - \ucb{R}_{-i}^{[k]}(s, a). 
% \end{equation}
%
% Note that the policy \( \ucb{q}^{[k+1]} \) evaluated in the second component on the right-hand side for \(\lcb{p}_{i}^{[k+1]}\) in Eq.~\eqref{eq:AlternativePaymentUpdate} is the same as the one in Eq.~\eqref{eq:PaymentUpdate-Payment} and comes from the policy that optimistically maximizes the social welfare calculated by using Eq.~\eqref{eq:AllocationPolicyUpdate} and implemented by the seller throughout episode \( (k+1) \). 
Note that the LCB $\lcb{R}_{-i}^{[k]}$ is used in Eq.~\eqref{eq:AlternativePaymentUpdate-Policy} and in the first component on the right-hand side of Eq~\eqref{eq:AlternativePaymentUpdate-Payment} whereas the UCB $\ucb{R}_{-i}^{[k]}$ is used in the second component on the right-hand side of Eq.~\eqref{eq:AlternativePaymentUpdate-Payment}.
In this scenario, each bidder \( i \in [n] \) tends to be slightly undercharged, resulting in a smaller amount of payment collected by the seller. Hence, this version is favorable to bidders but undesirable for the seller, and it can be adopted when the algorithm designer hopes to favor bidders.
 
In the next section, we present the main results of the version that is favorable to the seller. 
As for the alternative version favorable to bidders, our results on bounded regret, approximate efficiency, truthfulness, and individual rationality still hold (up to an additional factor polynomial in the size of the MDP, but the dependency on the time horizon or total number of episodes does not change), and the analysis is similar. 
Hence, the results and analysis for the alternative version will be omitted due to space limitation.

\section{Main Results}
\label{sec:results}

In this section, we present the main result of this paper, which provides a theoretical performance guarantee for our proposed learning algorithm (Algorithm \ref{alg:MainAlgorithm}), in terms of various regret metrics, as well as approximate efficiency, truthfulness, and individual rationality.
The proof of the main results can be found in Appendix \ref{ap:proof-main-results}.

\begin{thm}
\label{thm:OnlineLearning}
Given any $\epsilon\in (0,1)$, there exists a polynomially computable \( \d \in (0,1) \), such that with probability at least \( 1 - \Oh(\zeta) \), for \( T \geq \frac{8}{\a \d} \log ( A S / \zeta) \), the following results hold: 
\begin{enumerate}
    
    \item Suppose that all bidders are truthful. The social welfare regret satisfies 
    \begin{equation*}
        \Reg{SW}(T) \leq \wtl{\Oh} \left( \e T n + \a^{-\frac{4}{3}} \d^{-\frac{1}{3}} S^{-\frac{1}{2}} T^{\frac{2}{3}} n \right).
    \end{equation*}

    \item Suppose that all bidders are truthful. The seller's regret satisfies 
    \begin{equation*}
        \Reg{SELL}(T) \leq \wtl{\Oh} \left( \e T n^{2} + \a^{-\frac{4}{3}} \d^{-\frac{1}{3}} S^{-\frac{1}{2}} T^{\frac{2}{3}} n^{2} \right).
    \end{equation*}

    \item Suppose that all bidders are truthful. The bidders' regret satisfies 
    \begin{equation*}
        \Reg{BID}(T) \leq \wtl{\Oh} \left( \e T n^{2} + \a^{-\frac{4}{3}} \d^{-\frac{1}{3}} S^{-\frac{1}{2}} T^{\frac{2}{3}} n^{2} \right).
    \end{equation*}

    \item The learning algorithm is $\Oh(n \e)$-approximately efficient. 

    \item The learning algorithm is approximately individually rational. 

    \item The learning algorithm is approximately truthful.
\end{enumerate}
\end{thm}

\begin{rem}
When the time horizon $T$ is known to the seller a priori, by setting a proper value of $\delta$ so that $ \e \leq \Oh( T^{-1/3} ) $, the above three notions of regret are upper bounded by $ \wtl{\Oh} (T^{2/3}) $, which matches the upper bound $ \wtl{\Oh} (T^{2/3}) $ and the lower bound $ \Omega (T^{2/3}) $ for the MAB setting \cite{kandasamy2023vcg} and the episodic MDP setting \cite{qiu2024learning} up to a logarithmic factor. 
\end{rem}

\section{Conclusion}
\label{sec:conclusion}

In this paper, we study the problem of online learning for dynamic mechanism design in sequential auctions within an unknown environment. The problem is modeled as an infinite-horizon average-reward MDP, where the transition kernel and reward functions are unknown to the seller. We first extend the static VCG mechanism to the dynamic setting and obtain a dynamic VCG mechanism that preserves efficiency, truthfulness, and individual rationality. We then design an online learning algorithm for the seller to learn the dynamic VCG mechanism. The algorithm guarantees performance in terms of three notions of regret, upper bounded by $\wtl{\mathcal{O}} (\epsilon T + T^{2/3})$, and approximately satisfies efficiency, truthfulness, and individual rationality with high probability.
Some interesting directions for future work include relaxing Assumption~\ref{assump:NondegenerateMDP} and studying more general unichain or weakly communicating MDPs. Another promising extension is to develop a learning-based mechanism that is robust to bidders who use some learning dynamics, such as online gradient descent, to adaptively learn the mechanism.

% \begin{figure}
% \begin{center}
% \includegraphics[height=4cm]{jcaesar.eps}    % The printed column  
% \caption{Gaius Julius Caesar, 100--44 B.C.}  % width is 8.4 cm.
% \label{fig1}                                 % Size the figures 
% \end{center}                                 % accordingly.
% \end{figure}

% OR

%\begin{figure}
%\begin{center}
%\epsfig{file=jcaesar,width=7cm}
%\caption{Gaius Julius Caesar, 100--44 B.C.}
%\label{fig1}
%\end{center}
%\end{figure}

\bibliographystyle{plain}        % Include this if you use bibtex 
\bibliography{bibliography}      % and a bib file to produce the 
                                 % bibliography (preferred). The
                                 % correct style is generated by
                                 % Elsevier at the time of printing.

\appendix
% Each appendix must have a short title.
% Sections and subsections are supported  
% in the appendices.
% \newpage 

\section{Other Related Works}
\label{ap:related-works}

% \noindent
% \paragraph{Dynamic Mechanism Design}
\subsection*{Dynamic Mechanism Design}

There are extensive works on dynamic mechanism design.
Our work is closely related to \cite{bergemann2010dynamic} where the authors study a variant of MDP-based dynamic setting and generalize the static VCG mechanism to an infinite-horizon discounted-reward MDP. 
They propose a dynamic pivot mechanism that is efficient, ex post truthful and individual rational.
Earlier works that study MDP-based dynamic mechanisms include \cite{parkes2003mdp-based, parkes2004approximately, friedman2003pricing} where self-interested bidders arrive and depart over a finite time horizon and may misreport their types upon arrival.
\cite{pavan2014dynamic} generalize the Myerson mechanism to a dynamic environment and characterizes perfect Bayesian equilibrium-implementable allocation rules in Markovian settings. 
A more recent work on dynamic mechanism design in Markovian setting is \cite{zhang2022incentive} where each bidder observes her private state, chooses a reporting strategy and decides whether to stop immediately or to continue. 
Apart from the Markovian setting, \cite{gallien2006dynamic} studies dynamic mechanism design for selling multiple identical items to potential buyers that arrive over time.
\cite{kakade2013optimal} study settings where agents have evolving private information and designs an optimal virtual-pivot mechanism that satisfies ex post truthfulness and individual rationality. 
\cite{golrezaei2017auctions} study dynamic auctions from a different perspective where bidders are allowed to acquire additional information and refine their values at an additional cost. 
We refer the readers to the surveys \cite{vohra2012dynamic, bergemann2019dynamic} for more existing works on this topic.

\subsection*{Reinforcement Learning in Stochastic Games}

Our work is also related to the area of reinforcement learning in stochastic games for convergence to Nash equilibrium. 
Although in our setting, the problem is modeled as a single-agent MDP and reinforcement learning, the seller's allocation policy and payments are influenced by the bidding strategy of the bidders.
Therefore, our work has a flavor of multi-player games. 
Most existing works focus on two-player zero-sum stochastic games \cite{zhao2022provably, sayin2022fictitious, meila2021online}.
To tackle the problem of learning in multi-player stochastic games, \cite{uzzaman2020reinforcement} study reinforcement learning in mean-field games, and \cite{meigs2019learning} study online learning in aggregative games. These are the games with special structures. 
In \cite{etesami2024learning, qin2024learning}, the authors focus on decentralized reinforcement learning in $n$-player stochastic games with independent chains and show that the algorithm converge to a set of $\e$-Nash equilibrium in a weak sense.
We refer the readers to \cite{zhang2021multi, ozdaglar2021independent} for a broader review of the literature in this line of research. 

\section{Examples of Different Types of Auction}
\label{ap:examples}

\begin{exmp}[Single-item auctions]
    The seller allocates one item to one bidder or to nobody. The action space can be described as $\mc{A} = \{\mf{0}, \mf{e}_1, \cdots, \mf{e}_n\}$ where $\mf{0}$ is the all-zero vector, and $\mf{e}_i$ is an $n$-dimensional vector with the $i$-th entry being $1$ and the remaining entries being $0$. An action $\mf{e}_i$ represents allocating the item to bidder $i$, and the action $\mf{0}$ represents not allocating the item to anyone. 
    % The state space $\mc{S}$ can represent the set of knowledge or information about the item that the bidders have, and their knowledge or information evolves stochastically as the auction proceeds.
\end{exmp}
\begin{exmp}[Multi-unit auctions]
    The seller distributes $m$ identical items to $n$ bidders. The action space can be described as $\mc{A} = \{(a_1, \cdots, a_n) \in \mb{Z}^n_+ | \sum_{i=1}^n a_i \leq m\}$. An action $a \in \mc{A}$ is an $n$-dimensional non-negative integral vector where the $i$-th entry $a_i$ denotes the number of items allocated to bidder $i$.
    % Similar as the previous example, the state space $\mc{S}$ can represent the set of knowledge or information about the item that the bidders have, and their knowledge or information evolves stochastically as the auction proceeds.
\end{exmp}
\begin{exmp}[Combinatorial auctions]
    The seller distributes $m$ heterogeneous items to $n$ bidders. The action space can be described as $\actions = \{0, 1, \cdots, n\}^m$. An action $a \in \mc{A}$ is an $m$-dimensional integral vector where the $j$-th entry $a_j$ denotes the index of the bidder to whom the $j$-th item is assigned (if $a_j=0$, the $j$-th item is not assigned to anyone). 
    % Note that in this case , $|\mc{A}| = (n+1)^m$. 
    % The MDP model can in no way circumvent the computational issue associated with single-shot combinatorial auctions. 
    % Similarly, the state space $\mc{S}$ can represent the set of knowledge or information about all items that the bidders have. 
\end{exmp}
\section{Proof of Theorem \ref{thm:InfiniteHorizonMDPVCGMechanism}}
\label{ap:offline}

\paragraph*{Efficiency.}
When all bidders are truthful, efficiency is achieved, since the mechanism chooses the allocation policy that maximizes the average social welfare $ J(\pi; R) $, as shown in Eq. \eqref{eq:DynamicVCG-allocation}.

\paragraph*{Individual Rationality.} 
Fix an arbitrary bidder $i \in [n] $. 
When $i$ is truthful, her average utility $u_{i}$ is given by the following equation: 
\begin{align*}
    u_{i} & = J(\pi^{*}; r_{i} - p_i^*) \\
    & = J(\pi^{*}; R - J(\pi_{-i}^*; R_{-i})) \\
    & = J(\pi^{*}; R) - J(\pi_{-i}^*; R_{-i}) \\
    & = \max_{\pi \in \Pi} J(\pi; R) - \max_{\pi \in \Pi} J(\pi; R_{-i}) \\
    & \geq 0.
\end{align*}
The last inequality holds because $R \geq R_{-i} \geq 0$.

\paragraph*{Truthfulness.}
Fix an arbitrary bidder $i \in [n] $. 
When $i$ is truthful, from the proof for individual rationality, her average utility $u_i$ is given by the following equation:  
\begin{equation*}
    u_{i} = \max_{\pi \in \Pi} J(\pi; R) - J(\pi_{-i}^*; R_{-i}).
\end{equation*}
When $i$ is untruthful, she reports $b_{i}$ to the seller. The allocation policy $\pi'$ and the price policy for her $p'_i$ chosen by the mechanism are given by the following equations: 
\begin{align*}
    \pi' & \in \argmax_{\pi \in \Pi} \ J(\pi; b_{i} + R_{-i}), \\
    p'_i & = J(\pi_{-i}^*; R_{-i}) - R_{-i}.
\end{align*}
Hence, bidder $i$'s utility $\tl{u}_{i}$ when she is untruthful is as follows:
\begin{align*}
    \tl{u}_{i} & = J(\pi'; r_i - p'_i) \\
    & = J(\pi'; R - J(\pi_{-i}^*; R_{-i})) \\
    & = J(\pi'; R) - J(\pi_{-i}^*; R_{-i}).
\end{align*}
Therefore, 
\begin{equation*}
    \tl{u}_{i} - u_{i} = J(\pi'; R) - \max_{\pi \in \Pi} J(\pi; R) \leq 0.
\end{equation*}
\section{Proof of Lemmas in Section \ref{sec:online}}
\label{ap:online}

\subsection{Proof of Lemma \ref{lem:MixingTime}}

\begin{pf}
    First, it holds that 
    \begin{equation*}
    P^{\pi}(s'|s) = \sum_{a\in \actions} P(s' | s, a) \pi(a|s) \geq \a \sum_{a\in \actions} \pi(a|s) = \a.
    \end{equation*}
    Hence, each element of $P^{\pi}$ is lower bounded by $\a$, i.e., $P^{\pi} \geq \a \onematrix[S]{S}$.
    Define matrix $Q^{\pi} \triangleq \frac{1}{1-\a S} (P^{\pi} - \a \onematrix[S]{S})$.
    It can be verified that $Q^{\pi}$ is a row-stochastic matrix, i.e., $Q^{\pi} \geq 0$, and the sum of each row of $Q^{\pi}$ equals one. 
    Therefore, 
    \begin{align*}
        \| (\nu - \nu')^{\tp} P^{\pi} \|_{1} 
        & = \| (\nu - \nu')^{\tp} ( \a \onematrix[S]{S} + (1- \a S)Q^{\pi} ) \|_{1} \\ 
        & \mathop{=}^{\text{(a)}} \ (1 - \a S) \| (\nu - \nu')^{\tp} Q^{\pi} \|_{1} \\
        & \mathop{\leq}^{\text{(b)}} \ (1 - \a S) \| (Q^{\pi})^{\tp} \|_{1} \| \nu - \nu' \|_{1} \\
        & \mathop{=}^{\text{(c)}} \ (1 - \a S) \| (\nu - \nu')^{\tp} \|_{1},
    \end{align*}
    where equality (a) is due to the fact that $(\nu - \nu')^{\tp} \onematrix[S]{S} = 0$, inequality (b) is obtained by using the definition of matrix norms induced by vector norms, and equality (c) is obtained by using the result that for a column-stochastic matrix $(Q^{\pi})^{\tp}$, $\| (Q^{\pi})^{\tp} \|_{1} = \max_{1 \leq j \leq S} \sum_{i = 1}^{S} |Q^{\pi}_{ij}| = 1$.
\end{pf}

\subsection{Proof of Lemma \ref{lem:ApproxStationarity}}

\begin{pf}
    Denote by $\nu^{\tau^{[k]}}$ the one-step transition distribution at the first round of episode $k$.
    Conditioning on $\mc{F}^{[k-1]}$, we have $ \EXP{(\nu^{t})^{\tp}} = (\nu^{\tau^{[k]}})^{\tp} (P^{\pi^{[k]}})^{t - \tau^{[k]}}$.
    Moreover, due to stationarity, it holds that $(\nu^{[k]})^{\tp} P^{\pi^{[k]}} = (\nu^{[k]})^{\tp}$.
    Therefore, by applying Lemma \ref{lem:MixingTime} iteratively, we have 
    \begin{align*}
        \norm{\EXP{\nu^{t}} - \nu^{[k]}}_{1}
        & \leq (1 - \a S)^{t - \tau^{[k]}} \norm{\nu^{\tau^{[k]}} - \nu^{[k]}}_{1} \\
        & \stackrel{\text{(a)}}{\leq} 2 (1 - \a S)^{d^{[k]}} \\
        & = 2 (1 - \a S)^{\frac{\log{k}}{\a S}} \\
        & \stackrel{\text{(b)}}{\lesssim} 2 (1 - \a S)^{- \frac{\log{k}}{2 \log (1 - \a S)}} \\
        & = \frac{2}{\sqrt{k}},
    \end{align*}
    where inequality (a) is due to the fact that round $t$ is in the stationary phase of episode $k$ and that $\| \nu^{\tau^{[k]}} - \nu^{[k]} \|_{1} \leq 2$, and inequality (b) is from the result that for $x \approx 0$, $-\log (1 - x) \approx x$.
    To complete the proof, let $\r = \r^{t}$, $\r' = \r^{[k]}$, $\nu = \nu^{t}$, $\nu' = \nu^{[k]}$, and $\pi = \pi^{[k]}$.
    We have
    \begin{align*}
        \| \r - \r' \|_{1} 
        & = \sum_{\substack{s \in \states \\ a \in \actions}} |\r(s,a) - \r'(s,a)| \\
        & = \sum_{\substack{s \in \states \\ a \in \actions}} |\nu(s) \pi(a|s) - \nu'(s) \pi(a|s) | \\
        & = \sum_{\substack{s \in \states \\ a \in \actions}} |\nu(s) - \nu'(s)| \pi(a|s) \\
        & = \sum_{s \in \states} |\nu(s) - \nu'(s)| \\
        & = \| \nu - \nu' \|_{1}.
    \end{align*}
\end{pf}

% \noindent
\subsection{Proof of Lemma \ref{lem:SufficientExploration}}

\begin{pf}
    % Let $\mc{F}^{t} \triangleq \sigma( (s^{\tau}, a^{\tau}) : \tau \leq t)$ be the $\sigma$-algebra generated by the stochastic process $\{(s^{\tau}, a^{\tau})\}_{\tau \leq t}$ and $\mb{F} = (\mc{F}^{t})_{t \in \Z_{+}}$ be the filtration. 
    % By Lemma \ref{lem:StrictlyPositiveProbability}, it holds that for any $s \in \states$, $a \in \actions$ and $\mc{H} \in \mc{F}^{t-1}$, it holds that $\PR{(s^{t}, a^{t}) \neq (s, a) | \mc{H}} \leq 1 - \a \d$.
    % %
    % Fix $ k \in \Z_{++}$, and denote by $t_{k}$ the starting time of episode $k$. Let $h^{[k]} = d^{[k]} + l^{[k]}$ denote the total length of episode $k$.
    % Therefore, for any $s \in \states$ and $a \in \actions$, 
    %
    Define the event $E^{t} \triangleq \{(s^{t}, a^{t}) \neq (s, a)\}$. Then,
    \begin{align*}
        & \PR{\text{$ (s, a) $ is not visited in episode $ k $}} \\
        % \leq \ & \PR{\text{$ (s, a) $ is not visited in the stationary phase of episode $ k $}} \\
        \leq \ & \PR{E^{\tau^{[k]} + d^{[k]}} \cap \cdots \cap E^{\tau^{[k]} + d^{[k]} + l^{[k]} - 1}} \\
        \stackrel{\text{(a)}}{\leq} \ & (1 - \a \d)^{l^{[k]}} \\
        = \ & \left( 1 - \a \d \right)^{ \frac{ \max\{4, \sqrt{k} \} \log (ASk/\zeta) }{ \a \d } } \\
        % & \leq \left( 1 - \a \d^{2} \right)^{ \frac{ 2 \log (SAk/\zeta) }{ \alpha \delta^{2} } } \\
        \stackrel{\text{(b)}}{\lesssim} \ & \left( 1 - \a \d \right)^{- \frac{ 2 \log (ASk/\zeta) }{ \log (1- \a \d) } } \\
        \leq \ &  \frac{ \zeta }{ ASk^{2} },
    \end{align*}
    where inequality (a) is obtained by using Lemma \ref{lem:NoVisitProbabilityDecay}, and inequality (b) is from the result that for $x \approx 0$, $-\log (1 - x) \approx x$.
    Finally, taking a union bound over all $s \in \states$, $a \in \actions$ and $ k \in \Z_{++} $ gives the result.
\end{pf}

%%

% \noindent
\subsection{Proof of Lemma \ref{lem:ConfidenceSetForTransitionKernel}}

\begin{pf}
    The confidence radius $\e^{[k]}(s, a, s')$ follows a sample-dependent empirical Bernstein bound in \cite{maurer2009empirical}. Lemma \ref{lem:ConfidenceSetForTransitionKernel} is a result of Theorem 4 in \cite{maurer2009empirical} and follows the proof in \cite[Appendix B]{jin2020learning}.

    For $k = 0$, the statement trivially holds. 
    For $k \in \Z_{++}$, applying Theorem 4 in \cite{maurer2009empirical} and taking a union bound over all $k \in \Z_{++} $, $s, s' \in \states$, and $a \in \actions$, it holds that with probability at least $1 - 2 \pi^{2} \zeta / 3$, all $k \in \Z_{++} $, $s, s' \in \states$, and $a \in \actions$,
    \begin{align*}
    & \left| P(s'|s, a) - \mean{P}^{[k]}(s' | s, a) \right| \\
    \leq \ & \sqrt{\frac{ 2 \mean{P}^{[k]}(s' | s, a) (1 - \mean{P}^{[k]}(s' | s, a)) \log \left( {S^{2} A k^{2}}/{\zeta} \right)}{ \max \{1, N^{[k]}(s, a) - 1\}}} \\
        & + \frac{7 \log \left( { S^{2} A k^{2} }/{\zeta} \right)}{3 \max \{ 1, N^{[k]}(s, a) - 1 \}} \\
    \leq \ & 2 \sqrt{\frac{ \mean{P}^{[k]}(s' | s, a) \log \left( { S A k }/{\zeta} \right)}{\max \{ 1, N^{[k]}(s, a) - 1 \}}} \\
        & + \frac{14 \log \left( { S A k}/{\zeta} \right)}{3 \max\{1, N^{[k]}(s, a) - 1 \}} \\
    = \ & \e^{[k]}(s, a, s').
    \end{align*}
\end{pf}

%%

% \noindent
\subsection{Proof of Lemma \ref{lem:ConfidenceSetForRewardFunction}}

\begin{pf}
    This is a direct application of Azuma–Hoeffding inequality and union bound. 
\end{pf}

\section{Proof of Theorem \ref{thm:OnlineLearning}}
\label{ap:proof-main-results}

% \subsection*{Proof of Theorem \ref{thm:OnlineLearning}}

% In this subsection, we provide a sketch of proof for Theorem \ref{thm:OnlineLearning}.
% The full proof is included in Appendix \ref{ap:proof-main-results}

We first define some notations that will be used throughout the section. 
For all $s \in \states$ and $ a \in \actions $, let $\r^{t}(s, a) \triangleq \PR{(s^{t}, a^{t}) = (s, a)}$, and note that $\r^{t}$ is a valid but not stationary distribution over $\states \times \actions$.
Let $ \pi^{[k]} $ denote the policy induced by $ \ucb{q}^{[k]} $, and let $q^{[k]}$ denote the occupancy measure defined by $ \pi^{[k]} $ and the true transition kernel $P$.
Furthermore, let $ \r^{[k]} $, $ \ucb{\r}^{[k]} $, and $ \r^{*} $ be the occupancy measures in $\states \times \actions$ induced by $ q^{[k]} $, $ \ucb{q}^{[k]} $, and $q^{*}$, respectively.
Throughout the proof, we use $\r$ and $q$ interchangeably to represent an occupancy measure or a distribution. 
Let $\mc{F}^{[k-1]}$ be the $\sigma$-algebra adapted to all measurable events up to the end of episode $(k - 1)$ and $\mb{F} = \{ \mc{F}^{[k]} \}_{k \in \Z_{+}}$ be the filtration. 
The proof is divided into five parts as follows.  

%%

% \medskip
% \noindent
% \paragraph{Social Welfare Regret and Approximate Efficiency} 
\subsection*{Social Welfare Regret and Approximate Efficiency} 

Using occupancy measure, we can write expected social welfare regret as follows:
\begin{equation*}
    \Reg{SW}(T) = \sum_{t=1}^{T} \langle \rho^{*} - \EXP{\rho^{t}}, R \rangle,
\end{equation*}
where the expectation is taken with respect to the randomness of the MDP and of the algorithm.
Suppose that all bidders are truthful. In each round of the mixing phase, the regret incurred is at most $(n + c_{\max}) \leq (c_{\max} + 1) n$.
Therefore, the regret incurred in all mixing phases up to $T$ is trivially bounded by $ c_{0} n \sum_{k=1}^{K} d^{[k]} $, where $c_{0} =c_{\max} + 1$, and $K$ denotes the number of episodes up to $T$.

Next, we bound the regret incurred in the stationary phase.
Fix an arbitrary round $t \in [T]$ in the stationary phase of episode $k \in [K] $. 
The regret incurred in round $t$ is decomposed as follows:
% \todo{Use filtration. Condition on what??}
\begin{equation*}
\begin{aligned}
    \langle \rho^{*} - \EXP{\rho^{t}}, R \rangle = \ & \langle \rho^{*} - \rho_{\d}^{*}, R \rangle + \langle \rho_{\d}^{*} - \rho^{[k]}, R \rangle \\
    & + \langle \rho^{[k]} - \EXP{\rho^{t}}, R \rangle,
\end{aligned}
\end{equation*}
where $ \rho_{\d}^{*} \in \argmax_{\rho \in \D_{\d}(P)} \langle \rho, R \rangle $.
By Proposition \ref{prop:ShrunkPolytope}, the first term on the right-hand side is bounded as \( \langle \rho^{*} - \rho_{\d}^{*}, R \rangle \leq c_{0} n \e \). 
Conditioned on $\mc{F}^{[k-1]}$, by using Lemma \ref{lem:ApproxStationarity}, the last term is bounded as 
\(
    \langle \rho^{[k]} - \EXP{\rho^{t}}, R \rangle \leq \| \rho^{[k]} - \EXP{\rho^{t}} \|_{1} \| R \|_{\infty} \leq 2 c_{0} n / \sqrt{k}
\).
%
% By Proposition \ref{prop:ShrunkPolytope}, the first term on the right-hand side can be bounded as $ \left \langle \rho^{*} - \rho_{\d}^{*}, R \right \rangle \leq (n + c_{\max}) \e \leq c_{\max} n \e $. 
% By Assumption \ref{assump:ErgodicMDP} and the discussion in Subsection \ref{subsec:LearningInEpisodes}, the last term can be bounded as
% %
% %\begin{equation*}
% \(
%     \left \langle \rho^{[k]} - \rho^{t}, R \right \rangle \leq \norm{\rho^{[k]} - \rho^{t}}_{1} \norm{R}_{\infty} \leq 2 c_{\max} n / \sqrt{k}.
% \)
% %\end{equation*}
%
Conditioning on $\mc{F}^{[k-1]}$, to bound the second term, we further decompose it as follows:
\begin{align*}
    & \langle \rho_{\d}^{*} - \rho^{[k]}, R \rangle \\
    = \ \ & \underbrace{ \langle \r_{\d}^{*}, R - \ucb{R}^{[k-1]} \rangle}_{\text{$ \leq 0 $ when Lemma \ref{lem:ConfidenceSetForRewardFunction} holds}} 
        + \underbrace{ \langle \r_{\d}^{*} - \ucb{\r}^{[k]}, \ucb{R}^{[k-1]} \rangle}_{\text{$\leq 0$ by Eq.~\eqref{eq:AllocationPolicyUpdate}}} \\
    & + \langle \ucb{\r}^{[k]} - \r^{[k]}, \ucb{R}^{[k-1]} \rangle + \langle \r^{[k]}, \ucb{R}^{[k-1]} - R \rangle \\
    \mathop{\leq}^{\text{w.h.p.}} & c_{0} n \| \ucb{\rho}^{[k]} - \rho^{[k]} \|_{1} +  \| \ucb{R}^{[k-1]} - R \|_{\infty}.
    % & + \max_{(s, a)} | \ucb{R}^{[k-1]}(s, a) - R(s, a) |.
\end{align*}
The first term is upper bounded by using Lemma \ref{lem:OccupancyMeasureDifference2}, and the second term is upper bounded by using Lemma \ref{lem:SufficientExploration}, Lemma \ref{lem:ConfidenceSetForRewardFunction}, and Eqs.~\eqref{eq:EmpiricalRewardFunction}--\eqref{eq:ConfidenceRadiusForRewardFunction}.
% When Lemma \ref{lem:SufficientExploration} and \ref{lem:ConfidenceSetForRewardFunction} hold, we have $\left | \ucb{R}^{[k-1]}(s, a) - R(s, a) \right | \leq 2 \beta^{[k-1]}(s, a) \leq 4 \sqrt{{ \log (nSAk / \zeta) }/{k}}$ for all $(s, a) \in \states \times \actions$.
Putting everything together, we obtain the following upper bound of social welfare regret:
\begin{align*}
    & \Reg{SW}(T) \\
    = \ & \sum_{t=1}^{T} \langle \rho^{*} - \EXP{\rho^{t}}, R \rangle \\
    \mathop{\leq}^{\text{w.h.p.}} & c_{0} n \sum_{k=1}^{K} d^{[k]} 
    + c_{0} n \sum_{k=1}^{K} l^{[k]} \cdot \\
    & \left( \e + \frac{2}{\sqrt{k}} + \| \ucb{\rho}^{[k]} - \rho^{[k]} \|_{1} + 4 \sqrt{\frac{\log (A S k n/ \zeta) }{k}} \right).
    % & \leq c_{\max} n \sum_{k=1}^{K} d^{[k]} + c_{\max} n T \e 
    %     + c_{\max} n \sum_{k=1}^{K} l^{[k]} \left( 2 e^{-\frac{d^{[k]}}{\tau}} +  \norm{\ucb{\rho}^{[k]} - \rho^{[k]}}_{1} + 4 \sqrt{\frac{ \log (n A S k / \zeta) }{k}} \right). 
    % & \leq c_{\max} n \sum_{k=1}^{K} d^{[k]} + c_{\max} n T \e + 2 c_{\max} n \sum_{k=1}^{K} e^{-\frac{d^{[k]}}{\tau}} l^{[k]} \\
    % & \quad + c_{\max} n \sum_{k=1}^{K} l^{[k]} \norm{\ucb{\rho}^{[k]} - \rho^{[k]}}_{1} + 2 c_{\max} n \sum_{k=1}^{K} l^{[k]} \max_{(s, a) \in \states \times \actions} \beta^{[k]}(s, a).
\end{align*}

By using Lemma \ref{lem:ApproxStationarity} and Lemma \ref{lem:SufficientExploration} for \( d^{[k]} \) and \( l^{[k]} \), using Lemma \ref{lem:OccupancyMeasureDifference2} to bound \( \| \ucb{\rho}^{[k]} - \rho^{[k]} \|_{1} \), and using Lemma \ref{lem:NumberOfEpisodes} for $K$, the regret is thus bounded as follows: 
% Let $ d^{[k]} = \frac{1}{2} \tau \log k $ and $ l^{[k]} =  { \max\{4, \!\sqrt{k} \} \log (SAk/\zeta) }/{ \alpha \delta^{2}}.$ By applying Lemma \ref{lem:OccupancyMeasureDifference2} to $ \norm{\ucb{\rho}^{[k]} -\! \rho^{[k]}}_{1}$ and using Lemma \ref{lem:NumberOfEpisodes} to upper bound $K = \Oh(\a^{2/3} \d^{4/3} T^{2/3})$, we can get 
\begin{align*}
    & \Reg{SW}(T) \\
    \mathop{\leq}^{\text{w.h.p.}} & c_{0} n T \e + \frac{c_{0}}{\a S} n K \log K + \frac{c_{1}}{\a \d} n K \log (A K S / \zeta) \\
    & + \frac{c_{2}}{\a^{2} \d \sqrt{S} } n K \log^{\frac{3}{2}} (A K S / \zeta) \\
    & + \frac{c_{3}}{\a^{2} \d} n \sqrt{K} \log (A K S / \zeta) \\
    & + \frac{c_{4}}{\a \d} n K \log^{\frac{3}{2}} (A K S n / \zeta) \\
    \leq \ \ & c_{0} n T \e + c_{5} \a^{-\frac{4}{3}} \d^{-\frac{1}{3}} S^{-\frac{1}{2}} T^{\frac{2}{3}} n \log^{\frac{3}{2}} (A S T n / \zeta)\\
    & + c_{6} \a^{-\frac{5}{3}} \d^{-\frac{2}{3}} T^{\frac{1}{3}} n \log (A S T / \zeta)
\end{align*}
where $c_{0}$ up to $c_{6}$ are absolute constants.
Therefore, with probability at least \(1 - \Oh(\zeta)\), \( \lim_{T \to \infty} \frac{1}{T} \Reg{SW}(T) = \Oh(n \e) \), implying that Algorithm \ref{alg:MainAlgorithm} is $ \Oh ( n \e ) $-approximately efficient.

\subsection*{Seller's Regret}

By applying Lemma \ref{lem:SellersUtility}, we have
\(
    u_{0}(\pi^{*}, p^{*}) = - (n-1) \left \langle \rho^{*}, R \right \rangle + \sum_{i=1}^{n} \left \langle \rho_{-i}^{*}, R_{-i} \right \rangle
\).
% \todo{Not easy to see. Expand the expression, maybe write a short lemma in the appendix.}
% Suppose all bidders are truthful. 
By using a similar argument as in social welfare regret, the seller's regret incurred in the mixing phase is at most \( \frac{c_{0}}{\a S} n K \log K \). To bound his regret incurred in the stationary phase, fix an arbitrary round $t \in [T]$ in the stationary phase of episode $k \in [K] $. Conditioned on $\mc{F}^{[k-1]}$, the seller's expected utility in round $t$ is lower bounded as follows:
\begin{align*}
    \EXP{u_{0}^{t}} 
    = \ & \langle \EXP{\rho^{t}}, r_{0} \rangle + \sum_{i=1}^{n} \langle \ucb{\rho}_{-i}^{[k]}, \ucb{R}_{-i}^{[k-1]} \rangle \\
        & - \sum_{i=1}^{n} \langle \EXP{\rho^{t}}, \lcb{R}_{-i}^{[k-1]} \rangle \\
    \mathop{\geq}^{\text{(a)}}_{\text{w.h.p.}} & \langle \EXP{\rho^{t}}, r_{0} \rangle + \sum_{i=1}^{n} \left( \langle \ucb{\rho}_{-i}^{[k]}, {R}_{-i} \rangle - \langle \EXP{\rho^{t}}, R_{-i} \rangle \right) \\
    \mathop{=}^{\text{(b)}} \ \ & -(n-1) \langle \EXP{\rho^{t}}, R \rangle + \sum_{i=1}^{n} \langle \ucb{\rho}_{-i}^{[k]}, {R}_{-i} \rangle \\
    \mathop{\geq}^{\text{(c)}} \ \ & -(n-1) \langle \rho^{[k]} , R \rangle + \sum_{i=1}^{n} \langle \ucb{\rho}_{-i}^{[k]}, {R}_{-i} \rangle \\
    & - c_{1} n^{2}  \| \EXP{\r^{t}} - \r^{[k]} \|_{1},
\end{align*}
where inequality (a) is the result of Lemma \ref{lem:ConfidenceSetForRewardFunction} that $ \lcb{R}_{-i}^{[k-1]} \leq R_{-i} \leq \ucb{R}_{-i}^{[k-1]}$ with high probability, equality (b) is due to Lemma \ref{lem:SellersUtility}, and inequality (c) is the result of H\"{o}lder's inequality with $c_{1}$ being an absolute constant. 
Therefore, the seller's regret incurred in round $t$ can be upper-bounded as follows:
% the result of the bounds for $\norm{\r^{t} - \r^{[k]}}_{1}$ and $\norm{\ucb{\r}^{[k]} - \r^{[k]}}_{1}$, which are derived in the proof for social welfare regret,  
%
\begin{equation*}
\begin{aligned}
    \EXP{u_{0}(\pi^{*}, p^{*})\!-\! u_{0}^{t}} \mathop{\leq}^{\text{w.h.p.}} & (n-1) \underbrace{ \langle \rho^{[k]} \!-\! \rho^{*}, R \rangle}_{\text{$ \leq 0 $ by Eq.~\eqref{eq:DynamicVCGEquivalent-allocation}}} \\
    & + \sum_{i=1}^{n} \langle \rho_{-i}^{*} - \ucb{\rho}_{-i}^{[k]}, R_{-i} \rangle \\
    & + c_{1} n^{2}  \| \EXP{\r^{t}} - \r^{[k]} \|_{1}.
\end{aligned}
\end{equation*}
For each \( i \in [n] \), the inner product \( \langle \rho_{-i}^{*} - \widehat{\rho}_{-i}^{[k]}, R_{-i} \rangle \) has a form similar to the social welfare regret and can therefore be bounded in exactly the same way as was done for the social welfare regret. The last term on the right-hand side, \( c_{1} n^{2}  \| \EXP{\r^{t}} - \r^{[k]} \|_{1} \), can be bounded using Lemma~\ref{lem:ApproxStationarity}. Therefore, with probability at least \(1 - \Oh(\zeta)\), \( \Reg{SELL}(T) \leq c_{2} n \cdot \Reg{SW}(T) \) with \(c_{2}\) being an absolute constant, which implies \( \lim_{T \to \infty} \frac{1}{T} \Reg{SELL}(T) = \Oh(n^2 \e) \).

\subsection*{Bidders' Regret}

Suppose all bidders are truthful. 
Fix an arbitrary bidder $ i \in [n] $. 
Her regret incurred in each round is at most 1, so her regret incurred in the mixing phase is upper bounded by \( \frac{1}{\a S} K \log K \).
Next we bound her regret incurred in the stationary phase. 
Fix an arbitrary round $ t \in [T] $ in the stationary phase of episode $k \in [K]$.
% The average utility for bidder $i$ when the offline dynamic mechanism is implemented is upper bounded as follows:
% %
% \begin{align*}
%     u_{i}(\pi^{*}, p^{*}) 
%     & = \left \langle \r^{*}, R \right \rangle - \left \langle \r_{-i}^{*}, R_{-i} \right \rangle \\
%     & \stackrel{\textcircled{1}}{\leq} \left \langle \r^{*}, R \right \rangle - \left \langle \r_{-i}^{*}, \ucb{R}_{-i}^{[k-1]} \right \rangle + 2 c_{\max} n \max_{(s, a)} \beta^{[k-1]}(s, a) \\
%     & \stackrel{\textcircled{2}}{\leq} \left \langle \r^{*}, R \right \rangle - \left \langle \r_{-i}^{*}, \ucb{R}_{-i}^{[k-1]} \right \rangle + 2 c_{\max} n \max_{(s, a)} \beta^{[k-1]}(s, a)
% \end{align*}
% %
% where
% \begin{itemize}
%     \item \textcircled{1} is due to Lemma \ref{lem:ConfidenceSetForRewardFunction} that with high probability, $R_{-i} \geq \lcb{R}_{-i}^{[k-1]} \geq \ucb{R}_{-i}^{[k-1]} - 2c_{\max}n \beta^{[k+1]} $;
%     \item \textcircled{2}
% \end{itemize}
%%
By following the same steps,  conditioning on \( \mc{F}^{[k-1]} \), we can lower bound $i$'s expected utility in round $t$ as follows: 
\begin{align*}
    \EXP{u_{i}^{t}}
    = \ \ & \langle \EXP{\r^{t}}, r_{i} \rangle \\
        & - \left( \langle \ucb{\r}_{-i}^{[k]}, \ucb{R}_{-i}^{[k-1]} \rangle - \langle \EXP{\r^{t}}, \lcb{R}_{-i}^{[k-1]} \rangle \right) \\
    \mathop{\geq}^{\text{(a)}}_{\text{w.h.p.}} & \langle \EXP{\r^{t}}, R \rangle - \langle \ucb{\r}_{-i}^{[k]}, R_{-i} \rangle \\
        & - c_{1} n \max_{(s, a)} \beta^{[k-1]}(s, a) \\
    % & \geq \langle \EXP{\r^{t}}, r_{i} \rangle - \langle \r_{-i}^{[k]}, R_{-i} \rangle + \langle \r^{t}, R_{-i} \rangle \\
    % & \quad - c_{1} n \max_{(s, a)} \beta^{[k-1]}(s, a) - c_{2} n \norm{\ucb{\rho}^{[k]} - \rho^{[k]}}_{1} - c_{3} n \norm{\r^{t} - \r^{[k]}}_{1} \\
    \mathop{\geq}^{\text{(b)}} \ \ & \langle \EXP{\r^{t}}, R \rangle - \langle \r_{-i}^{[k]}, R_{-i} \rangle \\
        & - c_{1} n \max_{(s, a)} \beta^{[k-1]}(s, a) - c_{2} n \| \ucb{\rho}_{-i}^{[k]} - \rho_{-i}^{[k]} \|_{1},
\end{align*}
where inequality (a) is the result of Lemma \ref{lem:ConfidenceSetForRewardFunction}, and inequality (b) is the result of H\"{o}lder's inequality with $c_{1}$ and $ c_{2} $ being absolute constants.
Denote by $H$ the last two terms on the right-hand side of the last expression, which can be bounded by using Lemma \ref{lem:OccupancyMeasureDifference2} and Eq.~\eqref{eq:ConfidenceRadiusForRewardFunction}.
Therefore, bidder $i$'s regret incurred in round $t$ is upper bounded as follows: 
\begin{equation*}
\begin{aligned}
    & \EXP{u_{i}(\pi^{*}, p^{*}) - u_{i}^{t}} \\
    = \ \ & \langle \r^{*}, R \rangle - \langle \r_{-i}^{*}, R_{-i} \rangle - \EXP{u_{i}^{t}} \\
    \mathop{\leq}^{\text{w.h.p.}} & \langle \r^{*} - \EXP{\r^{t}}, R \rangle 
    + \underbrace{ \langle \r_{-i}^{[k]} -  \r_{-i}^{*}, R_{-i} \rangle}_{\text{$ \leq 0 $ by Eq.~\eqref{eq:DynamicVCGEquivalent-price1}}} - H.
\end{aligned}
\end{equation*}
The first term \( \langle \r^{*} - \EXP{\r^{t}}, R \rangle \) is exactly the social welfare regret incurred in round $t$.
Therefore, with probability at least \(1 - \Oh(\zeta)\), 
\(
    \Reg{BID}(T) \leq c_{3} n \cdot \Reg{SW}(T)
\) with \(c_{3}\) being an absolute constant, which implies that \( \lim_{T \to \infty} \frac{1}{T} \Reg{BID}(T) = \Oh(n^{2} \e) \).
% %
% \begin{align*}
%     \EXP{\Reg{BID}(T)} & \leq c_{4} n \cdot \EXP{\Reg{SW}(T)} \\
%     & \leq c_{5} n^{2} T \e + c_{6} \tau \a^{\frac{2}{3}} \d^{\frac{4}{3}} n^{2} T^{\frac{2}{3}} \log T+ c_{7} \a^{-\frac{1}{3}} \d^{-\frac{2}{3}} n^{2} |\mc{S}|^{\frac{1}{2}} T^{\frac{2}{3}} \log^{\frac{3}{2}} (n SAT / \zeta) \\
%     & \quad + c_{8} \a^{-\frac{1}{3}} \d^{-\frac{2}{3}} n^{2} |\mc{S}| T^{\frac{1}{3}} \log^{2} (SAT / \zeta),
% \end{align*}
% %
% where $c_{4}$ to $c_{8}$ are absolute constants. 
% This relation shows that $ \lim_{T \to \infty} \frac{1}{T} \EXP{\Reg{BID}(T)} = \Oh(n^2 \e) $.

% For approximate individual rationality, first, each bidder $i \in [n]$ is charged $0$ and thus always has nonnegative utility in the mixing phase.
% In the stationary phase, since $u_{i}(\pi^{*}, p^{*}) \geq 0$ when bidder $i$ reports truthfully (this is because the offline dynamic mechanism is individual rational), $ \lim_{T \to \infty} \frac{1}{T} \EXP{ \sum_{t=1}^{T} u_{i}^{t} } \geq \lim_{T \to \infty} \frac{1}{T} \EXP{ \sum_{t=1}^{T} (u_{i}^{t} - u_{i}(\pi^{*}, p^{*})) } \geq - \Oh( n \e ) $.
% Therefore, the learning algorithm is $\Oh( n \e )$-approximately individually rational. 

%%
\subsection*{Approximate Individual Rationality}

Fix an arbitrary bidder $ i \in [n] $. Suppose that bidder $i$ is truthful. 
In this part, we do not require that other bidders are truthful or adopt stationary bidding strategies. 
With slight abuse of notation, we still use $r_{j}^{t}$ to denote the reported reward of bidder $j \in [n] \setminus \{i\}$ in round $t$, and $\ucb{r}_{j}^{[k]}$ and $\lcb{r}_{j}^{[k]}$ to denote constructed UCB and LCB at the end of episode $k$.

Since bidder $i$ is charged $0$ in the mixing phase, her utility is nonnegative in the mixing phase. 
Therefore, we only need to lower bound her utility obtained in the stationary phase. 
Fix an arbitrary round $t \in [T]$ in the stationary phase of episode $k \in [K]$. Conditioned on $\mc{F}^{[k-1]}$, her expected utility in round $t$ is bounded as follows:
\begin{align*}
    \EXP{u_{i}^{t}}
    = \ \ & \langle \EXP{\r^{t}}, r_{i} \rangle \\
        & - \left( \langle \ucb{\r}_{-i}^{[k]}, \ucb{R}_{-i}^{[k-1]} \rangle - \langle \EXP{\r^{t}}, \lcb{R}_{-i}^{[k-1]} \rangle \right) \\
    \mathop{\geq}^{\text{(a)}}_{\text{w.h.p.}} & \langle \EXP{\r^{t}}, \ucb{R}^{[k-1]} \rangle - \langle \ucb{\r}_{-i}^{[k]}, \ucb{R}_{-i}^{[k-1]} \rangle \\
        & - c_{1} n \max_{(s, a)} \beta^{[k]}(s, a) \\
    \mathop{\geq}^{\text{(b)}} \ & \langle \ucb{\r}^{[k]}, \ucb{R}^{[k-1]} \rangle - \langle \ucb{\r}_{-i}^{[k]}, \ucb{R}_{-i}^{[k-1]} \rangle \\
        & - c_{1} n \max_{(s, a)} \beta^{[k-1]}(s, a) \\
        & - c_{2} n \left( \| \ucb{\rho}^{[k]} - \rho^{[k]} \|_{1} + \| \EXP{\r^{t}} - \r^{[k]} \|_{1} \right),
\end{align*}
where inequality (a) is the result of Lemma \ref{lem:ConfidenceSetForRewardFunction}, and inequality (b) is the result of H\"{o}lder's inequality, with $c_{1}$ and $c_{2}$ being absolute constants. 
The first two terms are lower bounded by $0$ by the following inequality:
% by using the definition of $\ucb{\r}^{[k]}$ and $\ucb{\r}_{-i}^{[k]}$ defined by Eqs. \eqref{eq:AllocationPolicyUpdate}--\eqref{eq:PaymentUpdate-Policy} and the fact that $\ucb{R}^{[k-1]} \geq \ucb{R}_{-i}^{[k-1]} \geq 0$ as follows: 
\begin{equation*}
\begin{aligned}
    & \langle \ucb{\r}^{[k]}, \ucb{R}^{[k-1]} \rangle - \langle \ucb{\r}_{-i}^{[k]}, \ucb{R}_{-i}^{[k-1]} \rangle \\
    = & \max_{q \in \D_{\d}(\P^{[k-1]})} \langle q, \ucb{R}^{[k-1]} \rangle - \max_{q \in \D_{\d}(\P^{[k-1]})} \langle q, \ucb{R}_{-i}^{[k-1]} \rangle \geq 0,
\end{aligned}
\end{equation*}
where the inequality holds because $\ucb{R}^{[k-1]} \geq \ucb{R}_{-i}^{[k-1]} \geq 0$.
The last two terms involving $c_{1}$ and $c_{2}$ are bounded in exactly the same way as we did earlier for regret.
% where inequality (a) is the result of an application of Lemma \ref{lem:ConfidenceSetForRewardFunction} to $r_{i}$ and the fact that $ \ucb{R}_{-i}^{[k-1]} - \lcb{R}_{-i}^{[k-1]} \leq 2 c_{\max} n \beta^{[k-1]}$ by Eqs. \eqref{eq:UCBForRewardFunction}--\eqref{eq:LCBForRewardFunction}. Moreover, inequality (b) is the result of H\"{o}lder's inequality. Finally, inequality (c) is by the definition of $\ucb{\r}^{[k]}$ and $\ucb{\r}_{-i}^{[k]}$ given in Eqs. \eqref{eq:AllocationPolicyUpdate}--\eqref{eq:PaymentUpdate-Policy}, and inequality (d) is due to the fact that $\ucb{R}^{[k-1]} \geq \ucb{R}_{-i}^{[k-1]} \geq 0$.
%
By following the same procedure as we did to bound the regrets, we have that with probability at least \(1 - \Oh(\zeta)\), \( \EXP{ \sum_{t=1}^{T} u_{i}^{t} } \geq - \wtl{\Oh}(T^{2/3}) \), and therefore, \( \lim_{T \to \infty} \frac{1}{T} \EXP{ \sum_{t=1}^{T} u_{i}^{t} } \geq 0 \). 
Therefore, the learning algorithm achieves approximate individual rationality.

\subsection*{Approximate Truthfulness}

Fix an arbitrary bidder $ i \in [n] $.
We consider two cases: when $i$ is truthful, and when $i$ is untruthful. 
We do not require that other bidders are truthful. 
Yet, we assume that all other bidders adopt stationary bidding strategies, and we fix their bidding strategies in the two cases.
Denote by $\tl{u}_{i}^{t}$ the utility of bidder $i$ in round $t$ when she is untruthful.
Similarly, we put ``$\sim$'' above a symbol to denote its counterpart when $i$ is untruthful.
In each round of the mixing phase, $\tl{u}_{i}^{t} - u_{i}^{t} \leq 1$. Hence, the total loss by bidding truthfully incurred in the mixing phase is trivially bounded by $\frac{1}{\a S} K \log K$.

Next, we bound the loss by being truthful incurred in the stationary phase. 
Fix a round $t \in [T]$ in the stationary phase of episode $k \in [K]$. When $i$ is truthful, conditioned on $\mc{F}^{[k-1]}$ and from the proof for approximate individual rationality, her expected utility obtained in round $t$ is lowered bounded as follows:
\begin{align*}
    \EXP{u_{i}^{t}} 
    \mathop{\geq}^{\text{(a)}}_{\text{w.h.p.}} & \langle \ucb{\r}^{[k]}, \ucb{R}^{[k-1]} \rangle - \langle \ucb{\r}_{-i}^{[k]}, \ucb{R}_{-i}^{[k-1]} \rangle \\
        & - c_{1} n \max_{(s, a)} \beta^{[k-1]}(s, a) \\
        & - c_{2} n \left( \| \ucb{\rho}^{[k]} - \rho^{[k]} \|_{1} + \| \EXP{\r^{t}} - \r^{[k]} \|_{1} \right) \\    
    % \langle \ucb{\r}^{[k]}, \ucb{R}^{[k-1]} \rangle - \langle \ucb{\r}_{-i}^{[k]}, \ucb{R}_{-i}^{[k-1]} \rangle - c_{1} n \max_{(s, a) \in \states \times \actions} \beta^{[k-1]}(s, a) \\
    % & - c_{2} \left( \| \ucb{\rho}^{[k]} - \rho^{[k]} \|_{1} + \| \EXP{\r^{t}} - \r^{[k]} \|_{1} \right) \\
    \mathop{\geq}^{\text{(b)}}_{\text{w.h.p.}} & \langle \ucb{\r}^{[k]}, \ucb{R}^{[k-1]} \rangle - \langle \ucb{\r}_{-i}^{[k]}, \ucb{\tl{R}}_{-i}^{[k-1]} \rangle \\
    & - c_{3} n \max_{(s, a)} \beta^{[k-1]}(s, a) \\
    & - c_{2} n \left( \| \ucb{\rho}^{[k]} - \rho^{[k]} \|_{1} + \| \EXP{\r^{t}} - \r^{[k]} \|_{1} \right) 
\end{align*}
where inequality (a) is directly obtained from the proof for approximate individual rationality, inequality (b) is the result of Lemma \ref{lem:ConfidenceSetForRewardFunction} and Eqs.~\eqref{eq:UCBForRewardFunction}--\eqref{eq:LCBForRewardFunction} that with high probability, $ \ucb{R}_{-i}^{[k-1]} - 2c_{\max} n \beta^{[k-1]} \leq R_{-i} \leq \ucb{\tl{R}}_{-i}^{[k-1]} $, and $c_1$ to $c_3$ are absolute constants. Denote by $G$ the last two terms involving $c_2$ and $c_3$ on the right-hand side of the last expression.

When $i$ is untruthful, the MDP follows a different trajectory, and as a result, the confidence sets for transition kernel and reward functions are different. 
This difference can be bounded by a small term since sufficient exploration is achieved, according to Lemma \ref{lem:SufficientExploration}.
% Following the same steps for proving approximate individual rationality, we can upper bound $i$'s expected utility in round $t$ when she is untruthful as follows: 
When $i$ is untruthful, conditioning on \(\mc{F}^{[k-1]}\), by following the same steps, we can upper bound her expected utility in round $t$ as follows:
\begin{align*}
    \EXP{\tl{u}_{i}^{t}} 
    \mathop{\leq}^{\text{w.h.p.}} & \langle \ucb{\tl{\r}}^{[k]}, \ucb{\tl{R}}^{[k-1]} \rangle - \langle \ucb{\tl{\r}}_{-i}^{[k]}, \ucb{\tl{R}}_{-i}^{[k-1]} \rangle \\
    & + c_{4} n \left( \| \ucb{\tl{\r}}^{[k]} - \tl{\r}^{[k]} \|_{1} + \| \EXP{\tl{\r}^{t}} - \tl{\r}^{[k]} \|_{1} \right) \\
    \mathop{\leq}^{\text{w.h.p.}} & \langle \ucb{\tl{\r}}^{[k]}, \ucb{R}^{[k-1]} \rangle - \langle \ucb{\tl{\r}}_{-i}^{[k]}, \ucb{\tl{R}}_{-i}^{[k-1]} \rangle \\
    & + c_{4} n \left( \| \ucb{\tl{\r}}^{[k]} - \tl{\r}^{[k]} \|_{1} + \| \EXP{\tl{\r}^{t}} - \tl{\r}^{[k]} \|_{1} \right) \\
    & + c_{5} n \max_{(s, a)} \beta^{[k-1]}(s, a)
\end{align*}
where $c_4$ and $c_5$ are absolute constants. Denote by $H$ the last two terms involving $c_4$ and $c_5$ on the right-hand side of the last expression.
Therefore, 
\begin{align*}
    \EXP{\tl{u}_{i}^{t} - u_{i}^{t}} 
    \mathop{\leq}^{\text{w.h.p.}} & \langle \ucb{\tl{\r}}^{[k]} - \ucb{\r}^{[k]}, \ucb{R}^{[k-1]} \rangle \\
    & + \langle \ucb{\r}_{-i}^{[k]} - \ucb{\tl{\r}}_{-i}^{[k]}, \ucb{\tl{R}}_{-i}^{[k-1]} \rangle - G + H.
    % + c_{2} \norm{\ucb{\rho}^{[k]} - \rho^{[k]}}_{1} + c_{3} \norm{\r^{t} - \r^{[k]}}_{1} \\
    % & + c_{1}' n \max_{(s, a)} \beta^{[k]}(s, a) + c_{4} \norm{\ucb{\tl{\r}}^{[k]} - \tl{\r}^{[k]}}_{1} + c_{5} \norm{\tl{\r}^{t} - \tl{\r}^{[k]}}_{1} + c_{6} n \max_{(s, a)} \tl{\beta}^{[k-1]}(s, a).
\end{align*}
Note that we cannot claim \(\langle \ucb{\tl{\r}}^{[k]} - \ucb{\r}^{[k]}, \ucb{R}^{[k-1]} \rangle \leq 0 \) or \( \langle \ucb{\r}_{-i}^{[k]} - \ucb{\tl{\r}}_{-i}^{[k]}, \ucb{\tl{R}}_{-i}^{[k-1]} \rangle \leq 0 \) by optimality like what we did for bounding the regrets. 
This is because $\ucb{\tl{\r}}^{[k]}$ is not necessarily in $\D_{\d}(\P^{[k-1]})$, and $\ucb{\r}_{-i}^{[k]}$ is not necessarily in $\D_{\d}(\tl{\P}^{[k-1]})$ either, since they are from two different trajectories. 
However, we can bound them by a small positive term as follows. 
Denote by $\tl{\pi}^{[k]}$ the policy induced by $\ucb{\tl{\r}}^{[k]}$, and let $\tl{\r}'^{[k]}$ be the occupancy measure defined by $P$ and $\tl{\pi}^{[k]}$.
According to Lemma  \ref{lem:ConfidenceSetForTransitionKernel}, $\tl{\r}'^{[k]} \in \D_{\d}(\P^{[k-1]})$ with high probability. 
Thus, it holds that 
\begin{align*}
    \langle \ucb{\tl{\r}}^{[k]} - \ucb{\r}^{[k]}, \ucb{R}^{[k-1]} \rangle
    \leq \ & \underbrace{ \langle \tl{\r}'^{[k]} - \ucb{\r}^{[k]}, \ucb{R}^{[k-1]} \rangle}_{\text{$ \leq 0 $ w.h.p. by Eq.~\eqref{eq:AllocationPolicyUpdate}}} \\
    & + c_{6} n \| \ucb{\tl{\r}}^{[k]} - \tl{\r}'^{[k]} \|_{1},
    % \mathop{\leq}^{\text{w.h.p.}} c_{6} n \| \ucb{\tl{\r}}^{[k]} - \tl{\r}'^{[k]} \|_{1},
\end{align*}
with $c_6$ being absolute constant, which can be further bounded by using Lemma \ref{lem:OccupancyMeasureDifference2}.
Similarly, with high probability, 
\(
\langle \ucb{\r}_{-i}^{[k]} - \ucb{\tl{\r}}_{-i}^{[k]}, \ucb{\tl{R}}_{-i}^{[k-1]} \rangle \leq c_{6} n \| \ucb{\r}_{-i}^{[k]} - \r'^{[k]}_{-i}\|_{1}
\),
where $\r'^{[k]}_{-i}$ is the occupancy measure defined by $P$ and the allocation policy induced by $\ucb{\r}_{-i}^{[k]}$. This term can be bounded by using Lemma \ref{lem:OccupancyMeasureDifference2}.

Putting everything together and following the same steps as we did for bounding the regrets, we have that with probability at least \( 1 - \Oh(\zeta) \), 
\( \EXP{ \sum_{t=1}^{T} (\tl{u}_{i}^{t} - u_{i}^{t})} \leq \wtl{\Oh}(T^{2/3}) \). 
Therefore, \( \lim_{T \to \infty} \frac{1}{T} \EXP{ \sum_{t=1}^{T} (\tilde{u}_{i}^{t} - u_{i}^{t})} \leq 0 \), which implies that the learned mechanism is approximately truthful.

\section{Auxiliary Lemmas and Proofs}
\label{ap:AuxiliaryLemmas}

\begin{lem}
\label{lem:StrictlyPositiveProbability}
    Let Assumption \ref{assump:NondegenerateMDP} hold.
    If $\ucb{q}^{[k]} \in \D_{\d}$ for all $k \in \Z_{++}$, then for all $t \in \Z_{+}$ and $t \geq 2$ and for all $s \in \states$ and $a \in \actions$, it holds that regardless of the history up to $t - 1$,
    \begin{equation*}
        \PR{ (s^{t}, a^{t}) = (s, a) } \geq \a \d.
    \end{equation*}
\end{lem}
\begin{pf}
    Fix an arbitrary round $t$ in episode $k$.
    If $\ucb{q}^{[k]} \in \D_{\d}$, by using Eq.~\eqref{eq:InducedPolicy}, it holds that for any $s \in \states$ and $a \in \actions$, $\pi^{[k]}(a | s) \geq \d$.
    If we condition on $(s^{t-1}, a^{t-1})$, for any $s' \in \states$ and $a' \in \actions$, it holds that 
    \(
        \PR{(s^{t}, a^{t}) = (s, a) | (s^{t-1}, a^{t-1}) = (s', a')} = \pi^{[k]}(a | s) P(s | s', a') \geq \alpha \delta
    \),
    independent of the history up to $t - 2$ by Markovian property.
    Taking the expectation of $(s^{t-1}, a^{t-1})$ gives $\PR{(s^{t}, a^{t}) = (s, a)} \geq \a \d$, regardless of the history up to $t - 1$.
\end{pf}

\begin{lem}
\label{lem:NoVisitProbabilityDecay}
    Let Assumption \ref{assump:NondegenerateMDP} hold.
    If $\ucb{q}^{[k]} \in \D_{\d}$ for all $k \in \Z_{++}$, then for all $t \in \Z_{+}$ and $t \geq 2$, for all $s \in \states$ and $a \in \actions$, and for any $l \in \Z_{+}$, it holds that regardless of the history up to $t - 1$, 
    \begin{equation*}
        \PR{ (s^{t}, a^{t}) \neq (s, a), \cdots, (s^{t+l}, a^{t+l}) \neq (s, a) } 
        \leq (1 - \a \d)^{l+1}.
    \end{equation*}
\end{lem}
\begin{pf}
    We prove the lemma using induction. 
    When $l = 0$, the statement is essentially Lemma \ref{lem:StrictlyPositiveProbability}.
    Suppose the statement holds for some $l \in \Z_{+}$. 
    Let $\mc{F}^{t+l} \triangleq \sigma( (s^{\tau}, a^{\tau}) : \tau \leq t + l)$ be the $\sigma$-algebra generated by the stochastic process $\{(s^{\tau}, a^{\tau})\}_{\tau \leq t + l}$. 
    By Lemma \ref{lem:StrictlyPositiveProbability}, we see that for all $s \in \states$, $a \in \actions$ and $\mc{H} \in \mc{F}^{t+l}$, it holds that $\PR{(s^{t+l+1}, a^{t+l+1}) \neq (s, a) | \mc{H}} \leq 1 - \a \d$.
    Define event $E^{\tau} \triangleq \{(s^{\tau}, a^{\tau}) \neq (s, a)\}$.
    We can see that $E^{t} \cap \cdots \cap E^{t + l} \in \mc{F}^{t+l}$.
    Therefore, 
    \begin{align*}
        \PR{E^{t} \cap \cdots \cap E^{t+l+1}} 
        & = \PR{E^{t+l+1} | E^{t} \cap \cdots \cap E^{t+l} } \cdot \\
            & \qquad \PR{E^{t} \cap \cdots \cap E^{t+l}} \\
        & \leq (1 - \a \d ) \cdot (1 - \a \d)^{l+1} \\
        & = (1 - \a \d)^{l+2}.
    \end{align*}
\end{pf}

\begin{lem}
\label{lem:OccupancyMeasureDifference1}
    For any $P' \in \P^{[0]}$, where $ \P^{[0]} $ is defined by Eq.~\eqref{eq:ConfidenceSetForTransitionKernelInitial} as follows:
    \begin{equation*}
    % \tag{\ref{eq:ConfidenceSetForTransitionKernelInitial}, revisited}
        \P^{[0]} \triangleq \set*{P \in \R_{+}^{A S^2} \given
        \sum_{s' \in \states} P(s' | s, a) = 1 \ \forall s \in \states, a \in \actions},
    \end{equation*}
    and any policy $\pi: \states \to \D(\actions)$, it holds that 
    \begin{equation*}
    \begin{aligned}
        & \norm{\r^{P, \pi} - \r^{P', \pi}}_{1} \\
        & \leq \frac{1}{\a S} \sum_{\substack{s \in \states \\ a \in \actions}} \r^{P', \pi}(s, a) \sum_{s'\in \states} \left| P(s' | s, a) - P'(s' | s, a) \right|.
    \end{aligned}
    \end{equation*}
\end{lem}
\begin{pf}
For simplicity, let $\r = \r^{P,\pi}$, $\r' = \r^{P', \pi}$, $\nu = \nu^{P,\pi}$, and $\nu' = \nu^{P', \pi}$. 
By using Lemma \ref{lem:ApproxStationarity}, we have \(\norm{\r - \r'}_{1} = \norm{\nu - \nu'}_{1} \).
% %
% \begin{align*}
%     \norm{\r - \r'}_{1} 
%     & = \sum_{\substack{s \in \states \\ a \in \actions}} |\r(s,a) - \r'(s,a)| \\
%     & = \sum_{\substack{s \in \states \\ a \in \actions}} |\nu(s) \pi(a|s) - \nu'(s) \pi(a|s) | \\
%     & = \sum_{\substack{s \in \states \\ a \in \actions}} |\nu(s) - \nu'(s)| \pi(a|s) \\
%     & = \sum_{s \in \states} |\nu(s) - \nu'(s)| = \norm{\nu - \nu'}_{1}.
% \end{align*}
%
Let $P(s' | s) = \sum_{a \in \actions} P(s' | s, a) \pi (a | s)$ and $P'(s' | s) = \sum_{a \in \actions} P'(s' | s, a) \pi (a | s)$ for all $s, s' \in \states$. 
We have 
\begin{align*}
    \norm{\nu - \nu'}_1 
    & = \norm{\nu^{\tp} P - (\nu')^{\tp} P'}_1 \\
    & = \norm{\nu^{\tp} P - (\nu')^{\tp} P + (\nu')^{\tp} P - (\nu')^{\tp} P'}_1 \\
    & \leq \norm{\nu^{\tp} P - (\nu')^{\tp} P}_1 + \norm{(\nu')^{\tp} P - (\nu')^{\tp} P'}_1 \\
    & \stackrel{\text{(a)}}{\leq} (1 - \a S) \norm{\nu - \nu'}_1 + \norm{(\nu')^{\tp} P - (\nu')^{\tp} P'}_1,
\end{align*}
where (a) is obtained by using Lemma \ref{lem:MixingTime}.
Therefore, 
\begin{align*}
    & \norm{\nu - \nu'}_1 \\
    \leq \ & \frac{1}{\a S} \norm{(\nu')^{\tp} P - (\nu')^{\tp} P'}_1 \\
    = \ & \frac{1}{\a S} \sum_{s'\in \states} \left| \sum_{s\in \states}\nu'(s) \left(P(s' | s) - P'(s' | s) \right) \right| \\
    = \ & \frac{1}{\a S} \sum_{s'\in \states} \Biggl| \sum_{s\in \states}\nu'(s) \cdot \\
    & \qquad \left(\sum_{a\in \actions} \pi(a|s) (P(s'|s,a) - P'(s'|s,a)) \right) \Biggr| \\
    \leq \ & \frac{1}{\a S} \sum_{s'\in \states} \sum_{s\in \states}\nu'(s) \sum_{a\in \actions} \pi(a|s) \cdot \\
    & \qquad \left| P(s'|s,a) - P'(s'|s,a) \right| \\
    = \ & \frac{1}{\a S} \sum_{\substack{s \in \states \\ a \in \actions}} \r'(s, a) \sum_{s'\in \states} \left| P(s' | s, a) - P'(s' | s, a) \right|.
\end{align*}
\end{pf}

\begin{lem}
\label{lem:OccupancyMeasureDifference2}
    With probability at least $1 - \Oh(\zeta)$, for all $k \in \Z_{++}$, it holds that
    \begin{equation*}
        \norm{\ucb{\r}^{[k]} - \r^{[k]}}_{1} \leq \frac{6}{\a \sqrt{S}} \sqrt{\frac{\log ({ASk}/{\zeta})}{\max\{1, k\}}} + \frac{20}{\a} \frac{ \log ({ASk}/{\zeta})}{\max\{1, k\}}.
    \end{equation*}
\end{lem}
\begin{pf}
    When Lemmas \ref{lem:SufficientExploration} and \ref{lem:ConfidenceSetForTransitionKernel} hold, for any $k \in \Z_{++}$, by applying Lemma \ref{lem:OccupancyMeasureDifference1}, we have 
    \begin{align*}
        & \norm{\ucb{\r}^{[k]} - \r^{[k]}}_{1} \\
        \leq & \frac{1}{\a S} \sum_{\substack{s \in \states \\ a \in \actions}} \ucb{\r}^{[k]}(s, a) \sum_{s'\in \states} \left| P(s' | s, a) - \ucb{P}(s' | s, a) \right| \\
        \leq & \frac{2}{\a S} \sum_{\substack{s \in \states \\ a \in \actions}} \ucb{\r}^{[k]}(s, a) \sum_{s'\in \states} \e^{[k]}(s, a, s') \\
        = & \frac{2}{\a S} \sum_{\substack{s \in \states \\ a \in \actions}} \ucb{\r}^{[k]}(s, a) 
        \sum_{s'\in \states} \Biggl( 2\sqrt{\frac{\widebar{P}^{[k]}(s'|s,a) \log ({ASk}/{\zeta})}{\max\{1, N^{[k]}(s, a)-1\}}} \\
        & \qquad + \frac{14 \log ({ASk}/{\zeta})}{3 \max\{1, N^{[k]}(s, a)-1\}} \Biggr) \\
        \leq & \frac{2}{\a S} \sum_{\substack{s \in \states \\ a \in \actions}} \ucb{\r}^{[k]}(s, a)
        \sum_{s'\in \states} \Biggl( 2\sqrt{\frac{\widebar{P}^{[k]}(s'|s,a) \log ({ASk}/{\zeta})}{\max\{1, k-1\}}} \\
        & \qquad + \frac{14 \log ({ASk}/{\zeta})}{3 \max\{1, k-1\}} \Biggr) \\
        \stackrel{\text{(a)}}{\leq} & \frac{2}{\a S} \sum_{\substack{s \in \states \\ a \in \actions}} \ucb{\r}^{[k]}(s, a) \Biggl( 2 \sqrt{\frac{S \log ({ASk}/{\zeta})}{\max\{1, k-1\}}} \\
        & \qquad + \frac{5 S \log ({ASk}/{\zeta})}{\max\{1, k-1\}} \Biggr) 
        \\
        = & \frac{4}{\a \sqrt{S}} \sqrt{\frac{\log ({ASk}/{\zeta})}{\max\{1, k-1\}}} + \frac{10}{\a} \frac{\log ({ASk}/{\zeta})}{\max\{1, k-1\}} \\
        \leq & \frac{6}{\a \sqrt{S}} \sqrt{\frac{\log ({ASk}/{\zeta})}{\max\{1, k\}}} + \frac{20}{\a} \frac{\log ({ASk}/{\zeta})}{\max\{1, k\}},
    \end{align*}
    where inequality (a) is due to Cauchy–Schwarz inequality. 
\end{pf}

\begin{lem}
\label{lem:NumberOfEpisodes}
    Let $T \in \Z_{++}$ and $ T \geq 4 \log ( AS / \zeta) / \a \d $. 
    Denote by $K$ the total number of episodes up to round $T$. 
    It holds that $
        K \leq \big \lceil \frac{3}{2} \left ( \a \d T \right)^{\frac{2}{3}} \big \rceil$.
\end{lem}
\begin{pf}
    It holds that 
    \begin{equation*}
    \begin{aligned}
        T & \geq \sum_{k = 1}^{K} \left( d^{[k]} + l^{[k]} \right) 
          \geq \sum_{k = 1}^{K} l^{[k]} \\
          & \geq \frac{1}{ \a \d } \sum_{k = 1}^{K} \sqrt{k} 
          \geq \frac{2}{ 3 \a \d } K^{\frac{3}{2}}.
    \end{aligned}
    \end{equation*}
\end{pf}

\begin{lem}
\label{lem:SellersUtility}
Given the offline infinite-horizon VCG mechanism that outputs \( (\pi^{*}, p^{*}) \), the seller's utility can be written as follows:
\begin{equation*}
    u_{0}(\pi^{*}, p^{*}) = - (n-1) \langle \rho^{*}, R \rangle + \sum_{i=1}^{n} \langle \rho_{-i}^{*}, R_{-i} \rangle.
\end{equation*}
\end{lem}

\begin{pf}
\begin{align*}
    u_{0}(\pi^{*}, p^{*}) 
    & = J(\pi^{*}; r_0 + \sum_{i=1}^{n} p_i^*) \\
    & = \langle \rho^{*}, r_0 + \sum_{i=1}^{n} p_i^* \rangle \\
    & = \langle \rho^{*}, r_0 - \sum_{i=1}^{n} R_{-i} \rangle + \sum_{i=1}^{n} \langle \rho^*_{-i}, R_{-i} \rangle \\
    % & = \langle \rho^{*}, R - \sum_{i=1}^n r_{i} \rangle + \sum_{i=1}^{n} \langle \rho^*_{-i} - \rho^*, R_{-i} \rangle \\
    & = \langle \rho^{*}, R - \sum_{i=1}^n (R_{-i} + r_{i})  \rangle + \sum_{i=1}^{n} \langle \rho_{-i}^{*}, R_{-i} \rangle \\
    & = - (n-1) \langle \rho^{*}, R \rangle + \sum_{i=1}^{n} \langle \rho_{-i}^{*}, R_{-i} \rangle.
\end{align*}
\end{pf}

\end{document}